\documentclass[12pt,a4paper]{article}
\usepackage{jheppub}

 \usepackage{graphicx}
  \usepackage{epstopdf}
  \usepackage{caption}
  \usepackage{subcaption}

\newcommand{\bej}[1]{ \begin{equation}\label{#1} }
\newcommand{\eej}{\end{equation}}
\newcommand{\beaj}[1]{\begin{eqnarray}\label{#1} }
\newcommand{\eeaj}{\end{eqnarray}}
\newcommand{\eq}[1]{(\ref{#1})}
\def\ZZZ{{\hskip-3pt\hbox{ Z\kern-1.6mm Z}}}
\def\zzz{{\hskip-3pt\hbox{ z\kern-1mm z}}}

\newcommand{\tp}{\tilde{\phi}}
\newcommand{\ttp}{\tilde{\tilde{\phi}}}
\newcommand{\tvp}{\tilde \varphi}

\newcommand{\vp}{\varphi}

\newcommand{\we}{\wedge}

\newcommand{\lm}{\lambda}

\newcommand{\half}{{1\over 2}}

\newcommand{\bd}{\bar{\rm D}}

\newcommand{\N}{\frac{m_{2}}{k_{2}}-\frac{m_{1}}{k_{1}}}

\newcommand{\be}{\begin{equation}}
\newcommand{\ee}{\end{equation}}
\newcommand{\ben}{\begin{eqnarray}\displaystyle}
\newcommand{\een}{\end{eqnarray}}

\newcommand{\p}{\partial}

\def\one{{\hbox{ 1\kern-.8mm l}}}
\def\zero{{\hbox{ 0\kern-1.5mm 0}}}

\def\be{\begin{equation}}       
\def\ee{\end{equation}}         

\def\bea{\begin{eqnarray}}      
\def\eea{\end{eqnarray}}
                                
\def\ba{\begin{array}}
\def\ea{\end{array}}
\def\bd{\begin{displaymath}}
\def\ed{\end{displaymath}}

\def\eq{\begin{equation}}
\def\eqe{\end{equation}}
\def\eqa{\begin{eqnarray}}
\def\eqae{\end{eqnarray}}
\def\ena{\end{eqnarray}}

\def\nn{\nonumber}

\def\unit{1 \hskip-.3em \raise2pt\hbox{$ \scriptstyle |$ } }

\def\a{\alpha}
\def\b{\beta}
 
\def\d{\delta}
\def\e{\epsilon}           
   
\def\g{\gamma}

\def\k{\kappa}                    
\def\l{\lambda}
\def\m{\mu}

  \def\w{\omega}
\def\p{\pi}                
  \def\th{\theta}                  
\def\s{\sigma}                                   
\def\t{\tau}

\def\D{\Delta}

\def\G{\Gamma}

\def\bd{\begin{displaymath}}
\def\ed{\end{displaymath}}
\def\quart{\frac14}
\def\6{\partial}
\def\N4{{\cal N}=4}

\def\half{{1 \over 2}}

\def\bop#1{\setbox0=\hbox{$#1M$}\mkern1.5mu
        \vbox{\hrule height0pt depth.04\ht0
        \hbox{\vrule width.04\ht0 height.9\ht0 \kern.9\ht0
        \vrule width.04\ht0}\hrule height.04\ht0}\mkern1.5mu}
\def\pa{\partial}                              

\def\>{\rangle} 

\def\<{\langle} 
\def\Dsl{D \hskip-.6em \raise1pt\hbox{$ / $ } }

\def\pa{\partial}

\def\+{\oplus}

\def\half{{1 \over 2}}

\title{Structure constants of $\beta$ deformed  super Yang-Mills}

\author{Justin R. David and } 
\author{Abhishake Sadhukhan}

\affiliation{ Centre for High Energy Physics,\\
Indian Institute of Science,\\ C.V. Raman Avenue, Bangalore 560012, India}

\emailAdd{justin, abhishake@cts.iisc.ernet.in}

\abstract{
We study the structure constants of the ${\cal N}=1$ beta 
deformed theory perturbatively  and at strong coupling.
We show that the planar one loop corrections to the 
structure constants of single trace gauge invariant operators 
in the scalar sector is  determined by the anomalous 
dimension Hamiltonian. 
This result implies that $3$ point functions of the chiral primaries of 
the theory do not receive corrections at one loop. 
We then study the structure constants 
at strong coupling using the Lunin-Maldacena geometry. 
We explicitly construct the supergravity mode dual to the chiral primary 
with  three equal $U(1)$ R-charges in the Lunin-Maldacena geometry. 
We show that the 3 point function of this
 supergravity mode  with semi-classical states 
representing two other similar chiral primary states
but with large $U(1)$ charges to be independent of the beta deformation and identical 
 to that found in the $AdS_5\times S^5$ geometry. 
This together with the one-loop result indicate that these 
 structure constants are protected by a non-renormalization theorem.
We also show that three point function of $U(1)$ R-currents with classical massive strings
is proportional to the R-charge carried by  the string solution. 
This  is in accordance with the prediction of the 
 R-symmetry Ward identity. }

\begin{document}
\maketitle
\section{Introduction}

Integrability has played a crucial role in determining the planar 
spectrum of anomalous dimensions
of gauge invariant single  trace operators in ${\cal N}=4$ Yang-Mills  theory at all 
orders in the t'Hooft coupling \cite{Beisert:2010jr}.  
The next piece of information one needs to solve a conformal field theory  are  
the fusion rules or the structure constants of the three point functions of gauge 
invariant operators. 
The presence of integrability in the structure constants 
of ${\cal N}=4$ Yang-Mills was first noticed in perturbative field theory calculations  in
\cite{Okuyama:2004bd,Roiban:2004va,Alday:2005nd}.
At strong coupling three point functions of chiral primaries was evaluated 
using supergravity in \cite{Lee:1998bxa}. 
These structure constants are not-renormalized and the fact that these also agree with the
the calculation at tree level  formed one of the early tests of the AdS/CFT correspondence. 
More recently methods to evaluate the structure constants at strong coupling involving one 
chiral primary and two other operators dual to classical spinning strings 
was developed in \cite{Janik:2010gc,Buchbinder:2010vw,Zarembo:2010rr,Costa:2010rz,Roiban:2010fe,Ryang:2010bn}. 
Finally in \cite{Escobedo:2010xs,Escobedo:2011xw,Foda:2011rr,Gromov:2012uv,Bissi:2012vx}
methods which use integrability of planar
${\cal N}=4$ Yang-Mills have been developed to evaluate structure constants. 
In  \cite{Escobedo:2011xw} 
the authors evaluated  the three point function of 
a chiral primary and two heavy (non-chiral) operators using integrability 
in a certain large charge limit and showed that this agreed with that 
obtained at strong coupling in 
\cite{Zarembo:2010rr}. 

Given the systematic progress achieved in  the study of three point functions in ${\cal N}=4$
Yang-Mills it is natural to ask if similar features exists for other integrable theories but with 
lower supersymmetry. One such theory which is a good candidate for such an exploration is
the beta deformed theory of Leigh and Strassler. 
Superconformal invariance of the beta deformed theory has been tested to 5 loops in
\cite{Elmetti:2006gr,Elmetti:2007up}. 
This theory is known to admit a holographic dual found by 
\cite{Lunin:2005jy}. 
Integrability in two point functions for the beta deformed theory has 
been observed in 
\cite{Frolov:2005iq,Beisert:2005if} 
and  the Y-system which in principle allows one to 
extract the anomalous dimensions of single trace operators has been written 
down in \cite{Gromov:2010dy,Fiamberti:2008sm,Arutyunov:2010gu,Ahn:2011xq}.
Integrability in three point functions of this theory has been not studied so far. 

In this paper  we study a class of the three point functions of ${\cal N}=1$ both from 
weak coupling and strong coupling. 
At weak coupling using the methods of 
\cite{Alday:2005nd} 
we show that the one loop corrections 
to the structure constants  of operators in the 
scalar sector are determined by the anomalous dimension Hamiltonian 
which is integrable. 
One of the implications of this result is that the structure constants of 
chiral primaries of this theory is not corrected at one loop.  
For arbitrary values of $\beta$ it is known \cite{Berenstein:2000hy,Berenstein:2000ux}
 that the chiral primaries of the
theory have the following $U(1)$ R-charges 
\begin{equation}
(k, 0, 0 ) , \qquad ( 0, k, 0) , \quad (0, 0, k), \quad (k, k, k). 
\end{equation}
The construction of the chiral primary with charge $(k, k, k)$ involves the deformation 
parameter $\beta$ \cite{Freedman:2005cg} \footnote{ A  perturbative study of BPS operators of this 
theory was also done in \cite{Penati:2005hp,Mauri:2005pa}.}.
However we show  the planar three point functions at tree level between
three such chiral primaries is independent of the beta deformation. Therefore at tree level
these structure constants are identical to that found for these states in ${\cal N}=4$ Yang-Mills. 
From our one loop calculation we  conclude  that  these
3-point functions do not get corrected as they involve chiral primaries and 
their value is identical to that of  ${\cal N}=4$ Yang-Mills  to one loop.

We then study  the structure constants at strong coupling using the Lunin-Maldacena
geometry.  We first construct the supergravity mode dual to the chiral 
primary with $U(1)$ R-charge $(k, k, k)$. 
We then use the method developed by \cite{Zarembo:2010rr} to evaluate the structure constant
of this supergravity mode with geodesics carrying  R-charge $(J, J, J)$. 
These geodesics are the semi-classical states dual to the chiral primary of 
interest. We find the structure constant to be independent of the 
coupling as well as independent of the beta deformation. 
Therefore even at strong coupling the structure constant involving  chiral primaries carrying 
equal R-charges  is identical to that of ${\cal N}=4$ Yang-Mills. 
This together with the one loop result suggests that  these structure 
constants are protected by a  non-renormalization theorem.

We also evaluate the structure constant of the R-currents of this theory with 
a generic massive semi classical string state and show that 
the structure constant is proportional to the R-charge carried by the
semi-classical solution. This is in accordance with that predicted by the
R-symmetry Ward identity. 
The verification of this  Ward identity at strong coupling 
for the case of ${\cal N}=4$  Yang-Mills  the 
verification of the Ward identity was  recently done in 
\cite{Georgiou:2013ff}. 
Finally we evaluate the structure constant of the supergravity mode with R-charge
$(k, k, k)$ and a rigid rotating string.

The organization of the paper is as follows:
In section 2 we  introduce the beta deformed ${\cal N}=1$ theory, this will help to set up notations
and conventions used in the paper. We  will also define what we mean by one loop 
corrections to structure constants. 
In section 3 we evaluate the  planar 
one loop  corrections to structure constants of gauge invariant
operators constructed out of the $3$ complex scalars and their conjugates. We show that the one loop 
correction is entirely determined by the anomalous dimension Hamiltonian of  the 
beta deformed theory. 
This allows us to conclude that the 3 point functions of chiral primaries do not 
get corrected at one loop. We then evaluate the the 3 point function of the 
chiral primary with charge $(k, k, k)$ and show that it is independent of the 
beta deformation and equal to that in the ${\cal N}=4$ Yang-Mills theory at one loop. 
In section 4 we turn to the evaluation of three point function at strong coupling. 
First we determine the supergravity mode  dual to the chiral primary 
with charge $(k, k, k)$
in the beta deformed gravity background. We use this mode 
to evaluate the 3 point function involving this chiral primary and geodesics carrying 
equal $U(1)$ charges. We show that that indeed that the 3 point function is 
independent of the beta deformation and identical to that 
of the ${\cal N}=4$ Yang-Mills. 
 We also evaluate three point functions involving $R$ currents and 
 massive semi-classical string states
show that these are proportional to the angular momentum of the 
semi-classical states as predicted by a conformal Ward identity. 
Finally we evaluate the structure constant of the supergravity mode dual to the 
$(k, k, k)$ chiral primary and the rigid rotating string. 
In appendix A we  review the evaluation of the one loop anomalous dimension of
the beta deformed theory.

\section{The beta deformed theory}

In this section we briefly review the beta deformed ${\cal N}=1$ Yang-Mills theory. 
This will serve to set up our  notations and conventions. 
The general deformation of $\mathcal{N}=4$ SYM which preserves $\mathcal{N}=1$ superconformal symmetry was first obtained by Leigh and Strassler \cite{Leigh:1995ep}. 
The field content of this theory is same as that of ${\cal N}=4$    Yang-Mills with 
gauge group $SU(N)$. It consists of a gauge field and its super partner, the gaugino. 
There are  $3$  complex scalars along with their super partners. 
All fields transform in the adjoint representation of  $SU(N)$.
 The  superpotential  of the theory is given by  
\begin{equation}
\label{superpot}
W=\kappa \left[{\mbox{Tr}}(e^{i\beta}\Phi_1\Phi_2\Phi_3-e^{-i \beta}\Phi_1\Phi_3\Phi_2)+\frac{h}{3}{\mbox{Tr}}(\Phi_1^3+\Phi_2^3+\Phi_3^3)\right],
\end{equation}
where parameters $\kappa$, $\beta$ and $h$ are complex. 
$\Phi_1, \Phi_2, \Phi_3$ are super fields containing the three complex scalars and
their super partners. 
The  beta deformed theory is a special case of the Leigh-Strassler deformation  
which is obtained by 
choosing $\kappa=g$,  where $g$ is the gauge coupling constant and 
$h=0$ with $\beta$ real.
This deformation is known to be a marginal  and the 
theory has ${\cal N}=1$ super-conformal symmetry.  
The  superpotential  in (\ref{superpot}) reduces to 
\be
W=g{\mbox{Tr}}(e^{i \beta}\Phi_1\Phi_2\Phi_3-e^{-i  \beta}\Phi_1\Phi_3\Phi_2).
\ee
The R-symmetry of this theory is $U(1)\times U(1) \times U(1)$. 
We now write down the explicit Lagrangian of the beta deformed theory
which we will use for all  our subsequent calculations \cite{Freedman:2005cg}
 
\begin{eqnarray}\label{betlag}
{\cal L}  &=&   {\rm Tr} \bigg(\quart F_{\mu \nu }^{~~2}+\half\bar{\lambda}
             {D}\!\!\!\!\slash\lambda
             +\overline{D_{\mu }Z^i }D_{\mu }Z^i +\half 
             \bar{\psi}^i{D}\!\!\!\!\slash\psi^i 
+ g (\bar{\lambda}\bar{Z}^i L\psi^i
  -\bar{\psi}^i R Z^i\lambda) \nonumber\\
 {}&& 
+\frac{1}{4} g^{2}[\bar{Z}^i, Z^i]^{2}\bigg)\nonumber\\ 
{}&& + \frac{g}{2}
{\rm Tr} \left(q\bar{\psi}_2 L Z_3\psi_1-\frac1q\bar{\psi}_1 L Z_3\psi_2\right)
+ \frac{g}{2}{\rm Tr} \left(\bar{q}\bar{\psi}_1 R
             \bar{Z}_3\psi_2-\frac{1}{\bar{q}}\bar{\psi}_2 R
             \bar{Z}_3\psi_1\right) \nn\\ 
{}&& +g^2{\rm Tr} \left((q Z_2 Z_3-\frac1qZ_3Z_2)(\bar{q} \bar{Z}_3
             \bar{Z}_2-\frac{1}{\bar{q}}\bar{Z}_2\bar{Z}_3)\right) \nn \\
 {}&& -\frac{g^2}{N} {\rm Tr} \left(q Z_2 Z_3-\frac1qZ_3Z_2\right) {\rm Tr}\left(\bar{q} \bar{Z}_3
             \bar{Z}_2-\frac{1}{\bar{q}}\bar{Z}_2\bar{Z}_3\right)+         {\rm cyclic}.
\end{eqnarray}
Note that there is a double trace operator in the last line of the Lagrangian given in (\ref{betlag})  
 for the $SU(N)$ theory. It is easy to see that in the planar limit and at one loop 
it affects only the anomalous dimensions and the structure constants 
of single trace operators involving only two scalars. For example, this term ensures that the operator ${\rm Tr}( Z_1 Z_2)$ does not receive corrections at planar one loop.  
We have also verified that the contribution of this term to the structure constants is
proportional to the anomalous dimension in accordance with the results of this paper.
For all other single trace operators
of length greater than two, the contribution of this term is supressed in the
large $N$ limit.  In this paper we will restrict our considerations
to single trace operators of length greater than two. Therefore from now on we will ignore this term in our analysis
\footnote{We thank Christoph Sieg for pointing out this subtelty to us \cite{Fokken:2013aea}.}.

The covariant derivatives are defined by 
\begin{eqnarray}
 D_{\m}Z_i &=&\6_{\m}Z_i-i\frac{g}{\sqrt{2}}[A_{\m},Z_i], \\ \nonumber
D_{\mu} \psi_i &=& \partial_{\mu} \psi_i - i \frac{g}{\sqrt{2}} [A_\mu, \psi_i], \\ \nonumber
D_\mu \lambda  &=& \partial_\mu\lambda - i \frac{g}{\sqrt{2}} [ A_\mu, \lambda].
\end{eqnarray} 
 Here the index  takes values in $i\in \{1,2,3\}$ and $q=e^{i\beta}$. 
 $L, R$ are the left and right chiral projectors. 
 The Lagrangian given in (\ref{betlag}) reduces to the 
${\cal N}=4$ Lagrangian if one chooses $\beta=0$.
The scalar potential plays an important role in our perturbative calculations, this is given by 
\begin{eqnarray}
\label{quad}
V&=&\frac{g^2}{4}{\mbox{Tr}}[\bar{Z_i}Z_i\bar{Z_j}Z_j+Z_i\bar{Z_i}Z_j\bar{Z_j}-\bar{Z_i}Z_iZ_j\bar{Z_j}-Z_i\bar{Z_i}\bar{Z_j}Z_j] \\ \nonumber
&&+g^2{\mbox{Tr}}[Z_2Z_3\bar{Z_3}\bar{Z_2}+Z_3Z_2\bar{Z_2}\bar{Z_3}-e^{i\beta}Z_2Z_3\bar{Z_2}\bar{Z_3}-e^{-i\beta}Z_3Z_2\bar{Z_3}\bar{Z_2}]+\mbox{cyclic}.
\end{eqnarray}
Here the first line contains D-type  terms of the potential while the second line contains the 
F-type terms.

We now discuss the observables of the beta deformed theory  we will be interested in this 
paper. Consider a basis of local  gauge invariant single trace operators such that their two point function
are diagonal and normalized in the  planar limit. 
That is 
\begin{equation}\label{2pt}
\langle {\cal O }_i (x_1) {\cal O}_j(x_2)  \rangle = \frac{\delta_{ij} }{| x_1-x_2|^{2\Delta_i} }.
\end{equation}
Here $\Delta_i$ is the conformal dimension of the operator $i$. In the planar limit 
it admits the following expansion in the t'Hooft coupling. 
\begin{equation}
\Delta= \Delta^{(0)} + \lambda \Delta^{(1)} + \lambda^2 \Delta^{(2))} + \cdots,
\end{equation}
where 
\begin{equation}
\lambda = g^2 N
\end{equation}
is the t'Hooft coupling. 
Once  we are in the above orthonormal basis of single trace local gauge invariant operator, 
the three point function  of any three operators is constrained by conformal 
invariance to be 
\begin{equation}\label{3pt}
\langle {\cal O}_i ( x_1)  {\cal O}_j(x_2)  {\cal O}_k(x_3)  \rangle =
\frac{C_{ijk}}{ |x_{12} |^{ \Delta_i + \Delta _j - \Delta _k} 
   |x_{23} |^{ \Delta_j + \Delta _k - \Delta _i} 
   |x_{31} |^{ \Delta_k + \Delta _i - \Delta _k}  },
   \end{equation}
   where $x_{ij} = |x_i - x_j|$. 
  $C_{ijk}$ are the structure constants of the theory.  It is easy to see form large $N$ counting
  that the leading  term in the  structure constants begin at $\frac{1}{N}$ in the large $N$ expansion. 
  Thus in the planar limit, the structure constants admit an expansion of the following 
  form
  \begin{equation}
   C_{ijk} = C_{ijk}^{(0)} + \lambda C_{ijk}^{(1)} + \lambda^2 C_{ijk}^{(2)} + \cdots
  \end{equation}
  In this paper we are interested in studying the properties of the structure  constants of the 
  beta deformed theory  in the planar limit both at the first order in the t'Hooft coupling as
  well as large t'Hooft coupling.

\section{Structure constants at one loop}
  
  We will now review the  method developed in \cite{Alday:2005nd} which we will use  obtain the planar structure constants at one loop in t'Hooft coupling. 
 The method captures the essential information necessary to evaluate the structure constant
 at one loop directly without constructing an orthonormal basis of operators at one loop. 
  Consider an arbitrary basis of gauge invariant operators constructed out of 
  the complex scalars. We denote these by  ${\cal O}_\alpha$, 
  this basis need not be orthonormal. 
  Then  their two point functions at one loop is given by 
  \begin{equation}
  \langle O_\alpha (x_1)  O_\beta (x_2) \rangle = 
  \frac{1}{|x_1 - x_2|^{2\Delta_\alpha} }\left( h_{\alpha\beta} + \lambda g_{\alpha\beta}
  - \lambda \gamma_{\alpha\beta} \ln ( ( x_1 -x_2) ^2 \Lambda^2 )  \right),
  \end{equation}
  where $\Lambda$ is the cut off. 
  Let $C^{(0)}_{\alpha\beta\gamma}$ be the tree level structure constants
  in this basis and $\tilde C^{(1)}_{\alpha\beta\gamma}$ be 
  finite part  of the one loop corrections to the  three point function defined by 
  \begin{eqnarray}
  \langle O_\alpha( x_1) O_\beta(x_2) O_\gamma(x_3) \rangle &=& 
  \frac{1}{|x_{12}|^{\Delta_\alpha + \Delta_\beta - \Delta_\gamma}}
   \frac{1}{|x_{23}|^{\Delta_\beta + \Delta_\gamma - \Delta_\alpha}}
    \frac{1}{|x_{31}|^{\Delta_\gamma + \Delta_\alpha - \Delta_\beta}} \times \\ \nonumber
    & & \left( C^{(0)}_{\alpha\beta\gamma} 
    - \lambda \gamma_{\alpha\delta} C^{(0) \delta}_{\,\beta\gamma} 
    \ln \left| \frac{x_{12} x_{13} \Lambda}{x_{23} } \right|
    - \lambda \gamma_{\beta\delta} C^{(0)\, \delta}_{\alpha\, \gamma}
     \ln \left| \frac{x_{12} x_{23} \Lambda}{x_{13} } \right| \right.  \\ \nonumber
    & & \left. - \lambda \gamma_{\gamma\delta} C^{(0) \,\, \delta}_{\alpha\beta}
     \ln \left| \frac{x_{13} x_{23} \Lambda}{x_{12} }\right| 
     + \lambda \tilde C^{(1)}_{\alpha\beta\gamma} \right).
  \end{eqnarray}
  Here
  \begin{eqnarray}
  & & C^{(0)\alpha}_{\, \beta\gamma} = h^{\alpha\delta}C^{(0)}_{\delta\beta\gamma}, 
  \quad
  C^{(0) \, \beta}_{\alpha\,\gamma} = h^{\beta\delta}C^{(0)}_{\alpha\delta\gamma}, 
  \quad
  C^{(0)\,\,\gamma}_{\alpha\beta} = h^{\gamma\delta} C^{(0)}_{\alpha\beta\delta}, \\ \nonumber
  & & h^{\alpha\gamma}h_{\gamma\beta} = \delta^{\alpha}_\beta.
  \end{eqnarray}
  The three point function is evaluated with the same regularization scheme as that used in the
  the two point function. 
  Then the renormalization group invariant 
   structure constants at one loop  is given by \cite{Alday:2005nd}
  \begin{equation}\label{reinvdef}
  C^{(1)}_{\alpha\beta\gamma} = \tilde C^{(1)}_{\alpha\beta\gamma} 
  -\frac{1}{2} g_{\alpha\delta} C^{(0)\delta}_{\, \beta\gamma} 
  -\frac{1}{2} g_{\beta\delta} C^{(0) \, \delta}_{\alpha\, \gamma}
  - \frac{1}{2} g_{\gamma\delta }C^{(0)\,\,\delta}_{\alpha\beta}.
  \end{equation}
Note that there is an additional  contribution to this structure constants 
at one loop when written out in the 
diagonal basis as defined in (\ref{2pt}) and (\ref{3pt}). 
This contribution arises due to the $O(\lambda)$ term in the mixing matrix on diagonalizing 
the two loop anomalous dimension Hamiltonian \cite{Georgiou:2009tp}. 
This added contribution
to the structure constants  will involve
the mixing matrix together with the  tree level structure constants. 
In the series of papers \cite{Georgiou:2012zj,Plefka:2012rd}  this problem has been addressed for 
${\cal N}=4$ Yang-Mills. In this paper we will restrict ourselves to studying 
the general properties  of the structure constants as defined in (\ref{reinvdef})
for the beta deformed theory.

\begin{figure}
\centering
\includegraphics[scale=1.0]{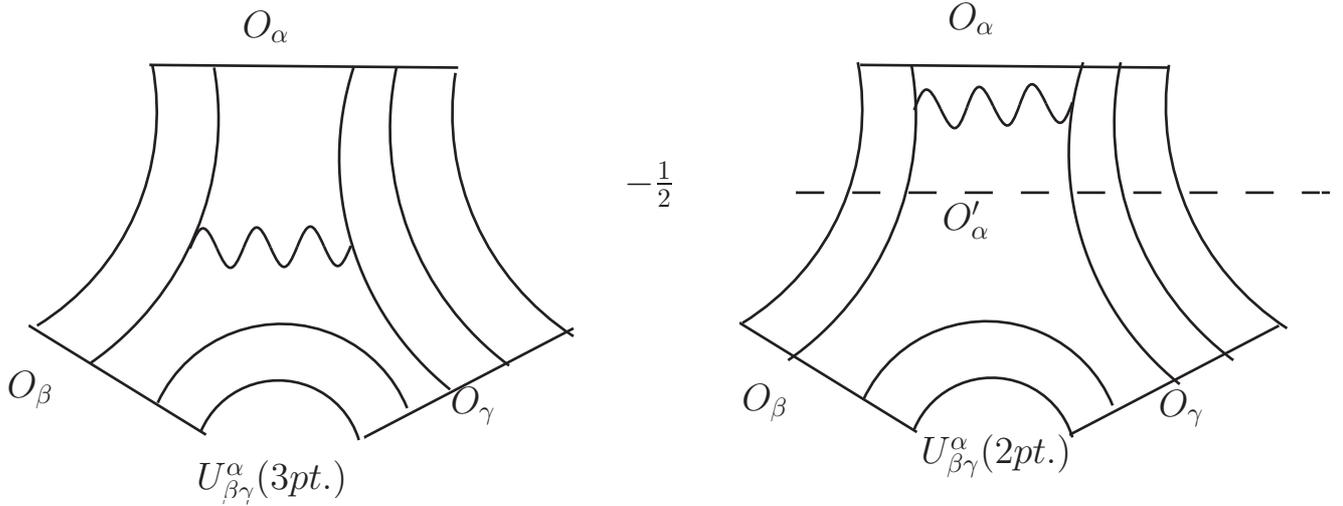}
\caption{Graph of renormalization scheme independent structure constant.}
\label{u32}
\end{figure}

  In \cite{Alday:2005nd} it was shown that an efficient way to evaluate this coefficient using
  planar  perturbative diagrams  is by using the formula
  \begin{eqnarray} \label{gen3pt}
  C^{(1)}_{\alpha\beta\gamma} &= &\left( 
  U^{\alpha}_{\beta\gamma}(3pt) - \frac{1}{2} U^\alpha_{\beta\gamma}(2pt) \right)
  + 
  \left( 
  U^{\beta}_{\gamma\a}(3pt) - \frac{1}{2} U^\beta_{\gamma\a}(2pt) \right) \\ \nonumber
  & &  + 
  \left( 
  U^{\g}_{\alpha\beta}(3pt) - \frac{1}{2} U^\g_{\alpha\beta}(2pt) \right).
  \end{eqnarray}
  Here $U^{\alpha}_{\beta\gamma} (3pt) $ refers to the finite terms in the 
   contribution of an interaction diagram that 
  has legs in all the three operators, with two legs in the nearest neighbour letters of 
  of operator ${\cal O}_\alpha$ and  one leg in operator ${\cal O}_\beta$ and 
  ${\cal O }_\gamma$ as shown in figure \ref{u32}.   $U^{\alpha}_{\beta\gamma}(2pt)$ 
  is the finite terms from the same diagram but now thought  of  a diagram 
  in a two point function. That is the letters from operators ${\cal O}_\beta$ and 
  ${\cal O }_\gamma$ are thought to be in the same operator. 
  The second diagram in figure \ref{u32} shows the operator these letters belong to by 
  the dashed line. 
  The rest of the contributions in (\ref{gen3pt}) are defined the same way. 
This formula for the renormalization scheme independent  one loop correction 
to the structure constant  does not depend on any  two body interaction diagrams, 
for example the self energy diagram (see figure(\ref{self})) between
any pairs of the operators. We can just focus on only the 3-body interaction diagrams.

We will now evaluate the one loop corrections to the structure constants as defined in 
  (\ref{gen3pt}) for scalar operators constructed out of only the three complex scalars in 
  the beta deformed theory.

\begin{figure}
\centering
\includegraphics[scale=1.0]{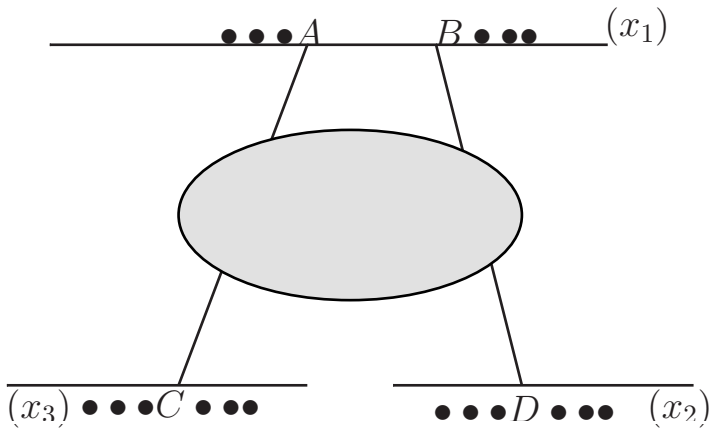}
\caption{A schematic one loop diagram}
\label{blob}
\end{figure}

\subsection{One loop diagrams and the contributions}

We will now explicitly evaluate the contribution of all possible diagrams at one loop that can 
contribute to the structure constants. 
As we have discussed earlier, the renormalization scheme independent 
contribution to the structure constants depend only on the combination given in 
(\ref{gen3pt}).   The gauge invariant single 
trace operators we consider consist of scalars built out of the 3 complex scalars of the 
beta deformed theory.  Thus we can restrict our selves to examine diagrams 
of the type given in figure \ref{blob}  with the letters $A, B, C, D$ belonging to any of the 
$3$ complex scalars and their conjugates.  The shaded blob represents either the quartic interactions 
due to the scalar potential  of (\ref{quad})
given in figures(\ref{x1}) and (\ref{x2}) or the gauge exchange given in figure(\ref{H}). 

After the evaluation of the contribution of each of the diagrams to the structure 
constants we will compare the contribution of the same diagram to 
the coefficient of the anomalous dimension matrix in a two point function. 
By this comparison we demonstrate that the contribution to the structure constants
by the same diagram is proportional to the anomalous dimension. 
This will establish that the contribution to the structure constants defined 
by (\ref{reinvdef}) is essentially captured by the anomalous dimension Hamiltonian. 

Note that by a simple large $N$ counting these diagrams are proportional to 
the factor $\frac{\lambda}{N}$ which we will suppress. 
Let us now examine the contribution of the diagrams to the structure constants
one by one. 

\subsubsection*{(i) $A = B = Z_i$ and  $C=D = \bar Z_i$, $i =1, 2, 3$ }

We consider the set of diagrams with both the letters $A, B$ being $Z_i$
and the letters $C, D$ being $\bar Z_i$.  $i$ can take any value $i=1, 2, 3$. 
The constant term from the quartic interaction for this diagram
is obtained by extracting the finite term
from 
\begin{equation}\label{q1}
 \tilde Q({\bf{i}}; {\rm 3pt} )  =-\frac{2}{4}  \left( \lim_{x_2 \rightarrow x_1} \frac{ \phi(r,s)}{x_{13}^2 x_{24}^2}
\right).
\end{equation}
Here the $2/4$ arises from the normalization of the quartic potential in the scalar potential 
given in (\ref{quad}). The quartic contribution arises from the figures (\ref{q3}) and (\ref{q4}).  $\phi(r, s)$ is given by the integral which 
appears  in the quartic interaction
\begin{equation}
 \int d^4 u \frac{1}{( x_1 - u)^2 (x_2 - u)^2 (x_3-u)^2 ( x_4-u)^2} = 
\frac{\pi^2 \phi(r, s)}{x_{13}^2 x_{24}^2} 
\end{equation}
and $r, s$ are the conformal cross ratios defined by 
\begin{equation}
 r = \frac{x_{12}^2 x_{34}^2}{x_{13}^2 x_{24}^2} , \qquad
s = \frac{x_{14}^2 x_{23}^2}{x_{13}^2 x_{24}^2} .
\end{equation}
The limit $x_2\rightarrow x_1$ is taken by setting 
$x_2 - x_1 = \epsilon \rightarrow 0$.  Note that under this limit
$r\rightarrow 0, s\rightarrow 1$.   The expansion of the function $\phi(r, s)$ is known
around this point \footnote{See equation B. 5 of \cite{Alday:2005nd}.}. 
Substituting this expansion  into (\ref{q1}) we obtain
\begin{equation}
 \tilde Q({\bf{i}}; {\rm 3pt} ) =  -\frac{2}{4} \frac{1}{x_{13}^2 x_{14}^2} \left( 
\ln\left[ \frac{x_{13}^2 x_{14}^2}{x_{34}^2 \epsilon^2} \right] + 2 \right).
\end{equation}
The term proportional to the logarithm contributes to the logarithmic
 corrections in the 3 point functions arising 
from the anomalous dimension, while the constant term contributes to the structure
constant.  Thus the constant term from the 3 body interaction of the quartic 
potential is given by 
\begin{equation}
 Q({\bf{i}}; {\rm 3pt} ) = -\frac{2}{4} \times 2.
\end{equation}
Now we examine the same diagram  but as a 2 body interaction. 
We thus have to evaluate
\begin{equation}
 \hat Q({\bf{i}}; {\rm 2pt} ) = -\frac{2}{4} \lim_{x_1\rightarrow x_2, x_3 \rightarrow x_4} 
\frac{ \phi(r,s)}{x_{13}^2 x_{24}^2},
\end{equation}
where the limits are taken  by setting 
$x_1\rightarrow x_2  =\epsilon$, $x_4- x_3=\epsilon$ and then finally 
taking $\epsilon\rightarrow 0$.  We can then extract the constant term from this using the 
expansions of the function $\phi(r, s)$. Performing  this we obtain
\begin{equation}
 Q( {\bf{i}};{\rm 2pt} ) = Q({\bf{i}};{\rm 3pt} )  = -1. 
\end{equation}

Let us now proceed to the  gauge exchange contribution to  the 3 body interaction. 
This is obtained by extracting the finite terms from 
\begin{equation}\label{lim3pt}
\hat G({\bf{i}}; {\rm 3pt} ) =  -\frac{1}{2} \lim_{x_1\rightarrow x_2} H(x_1, x_2, x_3, x_4) .
\end{equation}
Here the $1/2$ arises from the $\frac{1}{\sqrt{2}}$ in coupling in the covariant derivatives. 
The negative  sign  arises because two powers of $i$ in this interaction and $H$ is given by 
\begin{equation}
H = (\partial_1 - \partial_3) \cdot (\partial_2 -\partial_4) 
\int \frac{d^4 u d^4 v}{\pi^2 ( 2\pi)^2} \frac{1}{ (x_1 - u)^2 ( x_3 - u)^2 }
\frac{1}{(u-v)^2} \frac{1}{ (x_2 - v)^2 ( x_4 -v)^2 } .
\end{equation}
Setting $x_1 - x_2 =\epsilon$ and taking $\epsilon\rightarrow 0$ and keeping track of only the
finite terms we obtain \cite{Alday:2005nd}
\begin{equation}
G({\bf{i}}; {\rm 3pt} ) = -\frac{1}{2} \times 2.
\end{equation}
Note that as discussed in \cite{Alday:2005nd} there are terms in the limit (\ref{lim3pt})
which apparently seems to violate conformal invariance.  On summing 
all such terms from the contributions in the $3$ terms of (\ref{gen3pt}) it can be shown 
that they vanish. 
Let us now evaluate the contribution of the gauge exchange diagram but now 
thought of as arising from  a two point function. 
We have  extract the finite terms from 
\begin{equation}
 \hat G({\bf{i}}; {\rm 2pt}) = - \frac{1}{2} \lim_{x_1\rightarrow x_2, x_3\rightarrow x_4} 
H(x_1, x_2, x_3, x_4) .
\end{equation}
Evaluating this limit and extracting the finite terms  as done in \cite{Alday:2005nd}
we obtain
\begin{equation}
 G( {\bf{i}}; {\rm 2pt} ) = -\frac{1}{2} \times 6.
\end{equation}

Putting the contributions of the quartic interaction and the gauge exchange together 
and evaluating the renormalization scheme independent contribution to the 
structure constant we obtain
\begin{eqnarray}
C({\bf{i}}) &=& Q({\bf{i}}; {\rm 3pt} ) + H({\bf{i}}; {\rm 3pt} ) - \frac{1}{2} ( Q({\bf{i}}; ,{\rm 2pt}) + 
H({\bf{i}}; {\rm 2pt} )  ), \\ \nonumber
&=& 0.
\end{eqnarray}

Note that the contribution from the diagrams with 
$A=B= \bar Z_i$ and $C=D=\bar Z_i$, that is with the $Z$'s interchanged 
with the corresponding $\bar Z$ remains identical to the case discussed.

\subsubsection*{(ii) $ A = D = Z_i$ and $B=C= \bar Z_i$, $ i=1, 2, 3$}

Having explained in detail the contributions that make up diagram ({\bf i} ), we now just 
outline the results for the remaining diagrams. 
The contribution of  quartic interaction arises from figures (\ref{q1}) and (\ref{q2}) . This results in the 
following
\begin{equation}
 Q({\bf{ii}}; {\rm 3pt} ) -\frac{1}{2}  Q({\bf{ii}}; {\rm 2pt} ) = \frac{1}{2} .
\end{equation}
The gauge exchange diagram  gives
\begin{equation}
 G({\bf{ii}}; {\rm 3pt} ) -\frac{1}{2} G( {\bf{ii}}; {\rm 2pt} ) = \frac{1}{2} .
\end{equation}
Thus the total contribution of this diagram to the structure constant is 
\begin{equation}
 C({\bf{ii}}) = 1.
\end{equation}
As we have mentioned earlier, the same contribution arises from the complex 
conjugated diagram. 

\subsubsection*{(iii) $A=Z_i, B= Z_j, C= \bar Z_i, D= \bar Z_j$ and $i\neq j$, $i=1, 2, 3$}

The contribution of the quartic interaction here arises from both the F-term and 
D-terms in the scalar potential.  The D-term contribution arise from figures (\ref{q3}) and (\ref{q4}) and F contribution arise from (\ref{q6})(shown for the case of  $i=1,j=2$) and (\ref{q5})( shown
for the case of for $i=2,j=1$)
This results in the following
\begin{equation}
 Q({\bf{iii}}; {\rm 3pt} ) - \frac{1}{2} Q({\bf{iii}}; {\rm 2pt} ) = \frac{1}{2} .
\end{equation}
The gauge exchange diagram results in 
\begin{equation}
 G({\bf{iii}}; {\rm 3pt} ) -\frac{1}{2} G( {\bf{iii}}; {\rm 2pt} ) = \frac{1}{2}.
\end{equation}
Thus the total contribution to the structure constant is 
\begin{equation}
 C({\bf{iii}}) = 1.
\end{equation}
Again here the complex conjugate of this diagram gives the same contribution to the 
structure constant. 

\subsubsection*{(iv) $A= Z_i, B= \bar Z_j, C = \bar Z_i, D= Z_j$ and $i\neq j$, $i=1, 2, 3$}

The contribution  of the quartic interaction arises from diagram in figure(\ref{q2}). In fact this figure gives a factor of 2  due to the  two possible ways of Wick contraction in the  
D-terms. There is no contribution from the F-term. This results in the following
contribution to the structure constant
\begin{equation}
 Q({\bf{iv}}; {\rm 3pt} ) - \frac{1}{2} Q( {\bf{iv}}; {\rm 2pt} )  = \frac{1}{2} .
\end{equation}
The gauge exchange contribution results in 
\begin{equation}
  G({\bf{ii}}, {\rm 3pt} ) -\frac{1}{2} G( {\bf{ii}}, {\rm 2pt} ) = \frac{1}{2}.
\end{equation}
Thus the total contribution to the structure constant is
\begin{equation}
 C({\bf{iv}}) = 1.
\end{equation}
Again here the complex conjugate of this diagram results in same contribution to the 
structure constant. 

\subsubsection*{(v) $A= Z_i, B= \bar Z_i, C= \bar Z_j, D=  Z_j$ and $i\neq j$, $ i=1, 2, 3$}

For this case the  D-type terms of figures (\ref{q1})(with an extra factor of 2 for two ways of Wick contraction) contribute to the 
quartic  interaction. This results in 
\begin{equation}
 Q({\bf{v}}; {\rm 3pt} )- \frac{1}{2} Q( {\bf{v}}; {\rm 2pt} ) = \frac{1}{2} .
\end{equation}
There is no contribution from the gauge exchange interaction in this diagram. 
This is because one does not allow self contractions between the letters belonging
to the same operator. 
Therefore the total contribution is given by 
\begin{equation}
 C({\bf{v}}) = \frac{1}{2}.
\end{equation}
The complex conjugate of this diagram also yields the same result. 

\subsubsection*{(vi) $A =Z_i, B= \bar Z_i, C= Z_j, D= \bar Z_j$ and $i \neq j$, $ i = 1, 2, 3$}

D-type figures (\ref{q3}) and (\ref{q4}) contributes to the quartic interaction. F contribution arise from figure (\ref{q5})(for $i=1,j=2$) and (\ref{q6})(for $i=2,j=1$)
This results in 
\begin{equation}
 Q({\bf{vi}}; {\rm 3pt} ) - \frac{1}{2} Q({\bf{vi}}; {\rm 3pt} ) = \frac{1}{2}.
\end{equation}
Again for this diagram there is no contribution from the gauge exchange. 
Thus the structure constant contribution is 
\begin{equation}
 C({\bf{vi}}) = \frac{1}{2}.
\end{equation}
Here also the complex conjugate of this diagram yields the same result. 

\subsubsection*{(vii) $A = Z_1, B=Z_2, C= \bar Z_2 , D = \bar Z_1$  and cyclic}

Here only the F-type term contributes to the quartic interaction and there is 
no gauge exchange contribution. The diagram which contributes is (\ref{q7}). 
This results in the following
\begin{equation}
 C({\bf{vii}}) = Q( {\bf{vii}}; {\rm 3pt} ) - \frac{1}{2} Q( {\bf{vii}}; {\rm 2pt} ) = - e^{i\beta} .
\end{equation}
All diagrams related to this by the cyclic replacement of $Z_1 \rightarrow Z_2 \rightarrow Z_3 \rightarrow Z_1$ yields the same result. Further the complex conjugate of this diagram 
also yields the same result. 

\subsubsection*{(viii) $A =Z_2, B= Z_1, C= \bar Z_1 D= \bar Z_2$ and cyclic}

Only the F-type term given in figure (\ref{q8}) contributes, there is no contribution from the 
gauge exchange. This results in 
\begin{equation}
 C({\bf{vii}}) = Q( {\bf{viii}}; {\rm 3pt} ) - \frac{1}{2} Q({\bf{viii}}; {\rm 2pt}) = - e^{-i\beta} .
\end{equation}
Again all diagram related to this one by cyclic replacements and well as the 
conjugate diagram yields the same result for the structure constant. 

\subsubsection*{(ix) $A = Z_1, B = \bar Z_2, C = Z_2, D= \bar Z_1$ and cyclic}

The F-type term given in figure(\ref{q8}) contributes and there is no
contribution from the gauge exchange term. Thus the 
structure constant contribution from  this class of diagrams is given by 
\begin{equation}
C({\bf{ix}}) = Q( {\bf{ix}}; {\rm 3pt} ) - \frac{1}{2} Q( {\bf{ix}}; {\rm 2pt} ) = - e^{-i\beta}.
\end{equation}
All diagrams which are obtained by cyclic replacements as well as the conjugate 
diagrams result in the same structure constant. 

\subsubsection*{(x)   $A = Z_2, B= \bar Z_1, C= Z_1, D= \bar Z_2$ and cyclic}

\begin{figure}
 \centering
 \begin{subfigure}{.5\textwidth}
\centering
   \includegraphics[scale=0.8]{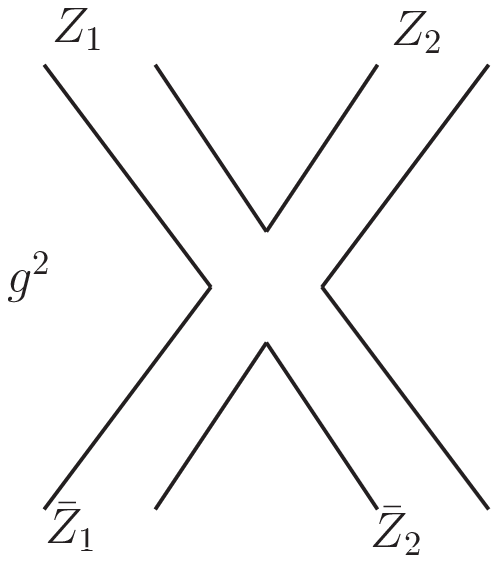}
   \caption{}
   \label{q5}
 \end{subfigure}%
 \begin{subfigure}{.5\textwidth}
   \centering
   \includegraphics[scale=0.8]{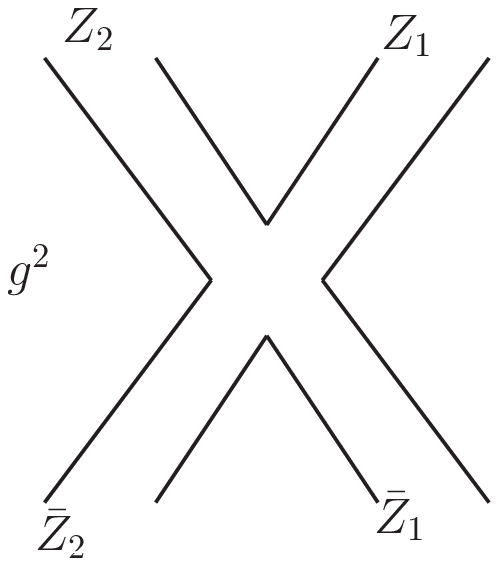}
   \caption{}
   \label{q6}
 \end{subfigure} 
\begin{subfigure}{.5\textwidth}
\centering
   \includegraphics[scale=0.8]{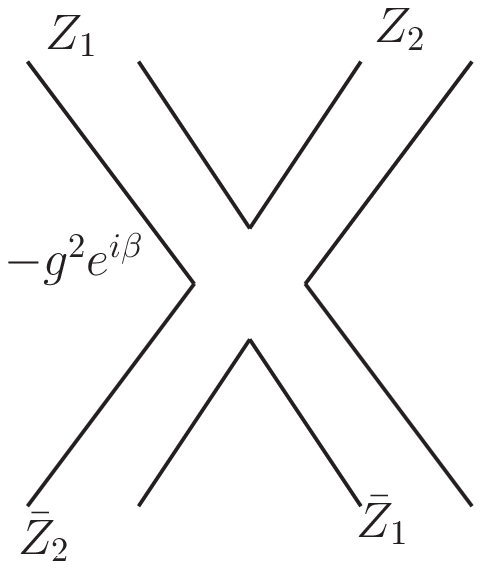}
   \caption{}
   \label{q7}
 \end{subfigure}%
 \begin{subfigure}{.5\textwidth}
   \centering
   \includegraphics[scale=0.8]{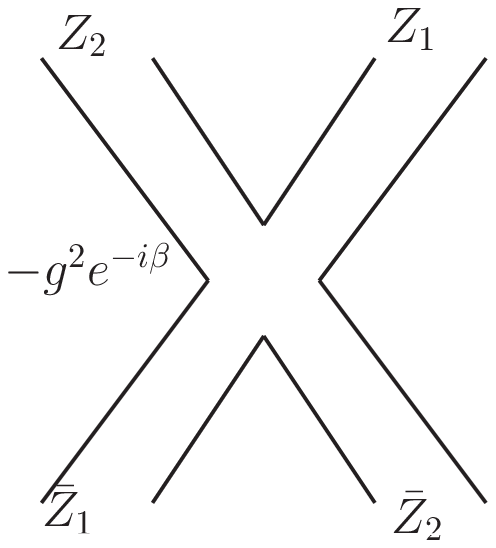}
   \caption{}
   \label{q8}
 \end{subfigure} 
 \caption{Possible F-type Quartic Interactions. Rotations of the diagrams in the plane are allowed in 
 Wick contractions. Cyclic
 replacements of $Z_1, Z_2$ are other possible diagrams}
\label{x2}
 \end{figure}

For these class of diagrams only the F-type term given in figure(\ref{q7}) contributes
and there is no contribution from the gauge exchange term. 
The contribution to the structure constant is given by 
\begin{equation}
C({\bf{x}}) = Q({\bf{x}}; {\rm 3pt} ) - \frac{1}{2} Q( {\bf{x}}; {\rm 2pt} )  = - e^{i\beta}.
\end{equation}
Again all diagrams obtained by cyclic replacements as well as the conjugate 
diagrams result in the same structure constant.

Now we can compare the evaluation of the contributions to the renormalization 
scheme independent structure constant to that of the anomalous
dimension Hamiltonian  given in the appendix A diagram by diagram . 
We see that the contribution to the structure constant for each diagram is 
$-1/2$ that  of  the anomalous dimension Hamiltonian. 
This shows that the one loop corrections to structure constants are indeed controlled 
by the anomalous dimension Hamiltonian. 
When all the contributions to the structure 
constant in given three point functions are put together these will organize into combinations of
the 
anomalous dimensions of the respective operators involved in the three point function. 
This is identical to how the coefficients of the logarithms in the three point function 
organize into combinations of the anomalous dimensions of the respective operators. 

\begin{figure}
 \centering
 \begin{subfigure}{.5\textwidth}
\centering
   \includegraphics[scale=0.7]{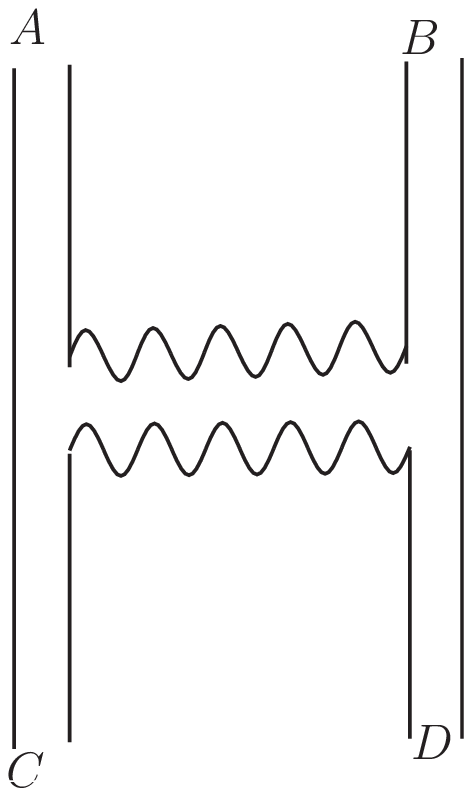}
   \caption{Gauge interaction}
   \label{H}
 \end{subfigure}%
 \begin{subfigure}{.5\textwidth}
   \centering
   \includegraphics[scale=0.7]{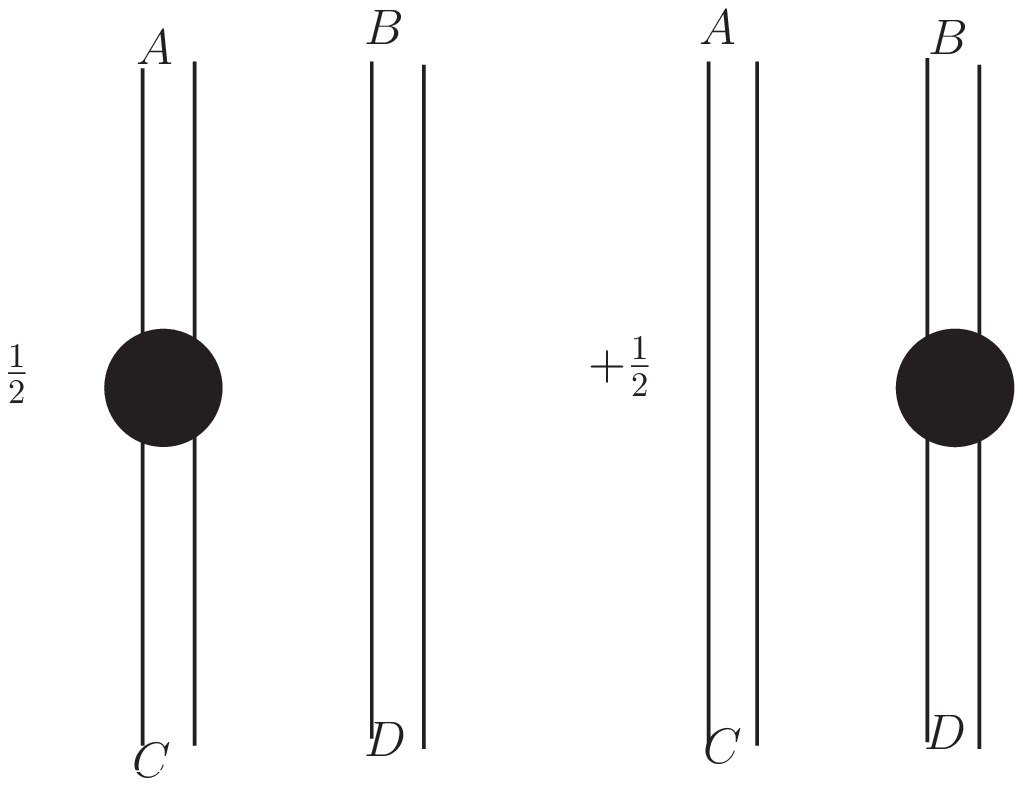}
   \caption{Self energy: fermions and bosons in loops}
   \label{self}
 \end{subfigure} 
\caption{Gauge exchanges and self energies}
\label{hs}
\end{figure}
 
 \begin{figure}[t]
 \centering
 \begin{subfigure}{.5\textwidth}
\centering
   \includegraphics[scale=0.8]{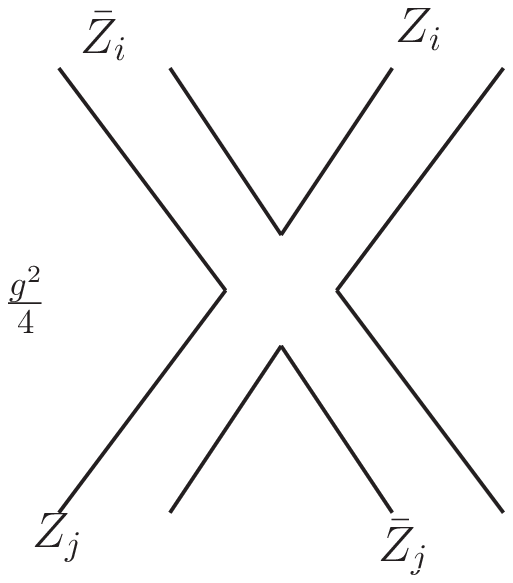}
   \caption{}
   \label{q1}
 \end{subfigure}%
 \begin{subfigure}{.5\textwidth}
   \centering
   \includegraphics[scale=0.8]{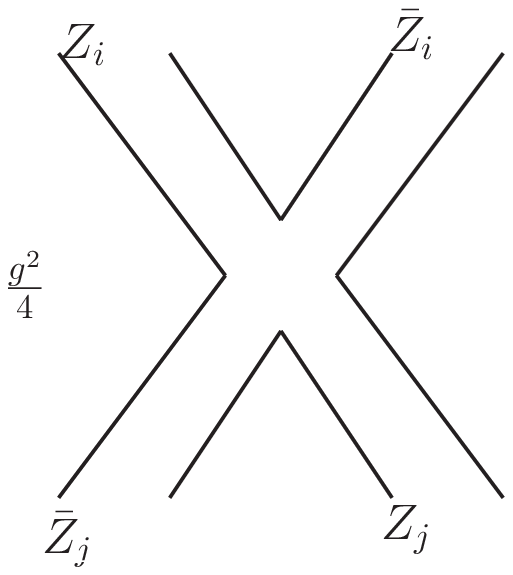}
   \caption{}
   \label{q2}
 \end{subfigure} 
\begin{subfigure}{.5\textwidth}
\centering
   \includegraphics[scale=0.8]{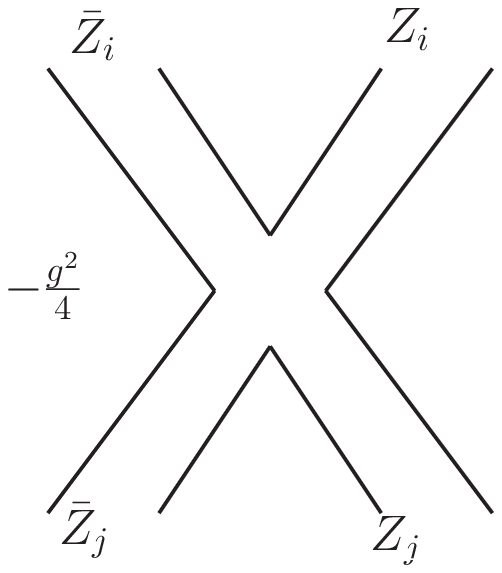}
   \caption{}
   \label{q3}
 \end{subfigure}%
 \begin{subfigure}{.5\textwidth}
   \centering
   \includegraphics[scale=0.8]{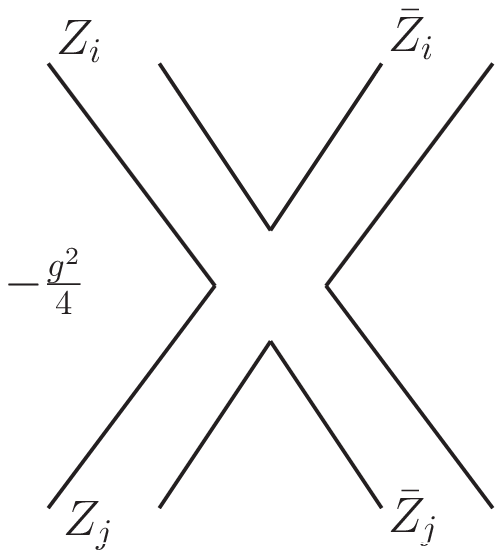}
   \caption{}
   \label{q4}
 \end{subfigure} 
 \caption{Possible D-type Quartic Interactions. Rotations of the diagrams in the plane are  allowed
 in Wick contractions}
\label{x1}
 \end{figure}

\subsection{Non-renormalization of structure constants of chiral primaries}

It is known  that the structure constants of $3$ chrial primaries 
in ${\cal N}=4$ Yang-Mills do not get renormalized at all orders in 
the coupling \cite{Lee:1998bxa}.  For a proof involving ${\cal N}=4$ harmonic
superspace see \cite{Howe:1998zi}\footnote{See 
\cite{Basu:2004nt,Baggio:2012rr} for recent proofs. }.
A similar question  for  the case of  the $\beta$ deformed ${\cal N}=1$ Yang-Mills is 
yet to be investigated. 
For arbitrary values of the deformation parameter $\beta$, the chiral primaries
have the following $U(1)$ charges
\begin{equation}\label{cp2}
(k, 0, 0), \qquad (0, k, 0), \qquad (0, 0, k), \qquad (k, k, k)
\end{equation}
 An argument to indicate that the three point function 
involving chiral primaries only of the first three type are 
not renormalized at one loop was given in \cite{Freedman:2005cg}.

Let us now use the result of the previous section to show  how 
three point function involving any  chiral primaries in (\ref{cp2}) including that 
with $U(1)$ charge $(k, k, k)$ are not renormalized at one loop in 
the t'Hooft limit.   We have shown that the finite corrections to structure constants
at one loop organized themselves to the anomalous dimensions of the 
operators involved in the three point function. If the three operators involved are
chiral primaries then this correction vanishes. 
As mentioned in the discussion below equation (\ref{reinvdef}), 
there is an additional  contribution to the structure constant
at one loop. This arises if the operators involved mix at one loop which can be 
found by examining the two loop anomalous dimension Hamiltonian. 
If the operators 
are all chiral primary, there is no mixing and hence this contribution also vanishes. 
Thus there are are no corrections at one loop to the structure constants 
involving only chiral primaries.

Let us examine the chiral primary with charge $(k, k, k)$ in more detail.  
The explicit construction of this  operator in the planar limit is 
by  \cite{Freedman:2005cg}
\begin{equation}\label{defcp}
 { O}_{(k, k, k)}  = \sum_\pi  c_\pi{\rm Tr} \left( \pi Z_1^k Z_2^k Z_3^k \right)  ,
\end{equation}
where 
\begin{equation}
 c_\pi = \frac{  \bar q ^{k_\pi}  }{s_\pi} , \qquad \bar q = e^{-i\beta}. 
\end{equation}
 $k_\pi$ is the number of permutations required to obtain that term from the 
configuration of 
\begin{equation}
 {\rm Tr} ( Z_1^k Z_2^k Z_3^k).
\end{equation}
$s_\pi$ is the symmetry factor which counts the number of  repeated  arrays, see
\cite{Freedman:2005cg} for a detailed discussion of how this  symmetry factor is evaluated. 
The sum in (\ref{defcp}) runs over all the permutations. 
Note that when $\beta=0$, it reduces to the chiral primary of ${\cal N}=4$. 
Let us explicitly write down the operators at the first two levels 
\footnote{The anomalous dimension of the second operator
for the case of the 
 general  Leigh-Strassler deformation was studied in \cite{Madhu:2007ew}}. 
\begin{eqnarray}\label{exam}
{O}_{(1, 1, 1)} &=& {\rm Tr}( Z_1 Z_2 Z_3) + \bar q^2 {\rm Tr}( Z_1 Z_2 Z_3) , 
\\ \nonumber
{O}_{(2,2,2)} &=& {\rm Tr} \left( Z_1^2Z_2^2 Z_3^2 + \bar q^2 Z_1^2 Z_2 Z_3Z_2Z_3 
+ \bar q^4 Z_1^2 Z_2 Z_3 ^2 Z_2 + Z_1^2 Z_3 Z_2^2 Z_3 \right. \\ \nonumber
& & \bar q^6 Z_1^2 Z_3Z_2 Z_3 Z_2 + \bar q^8 Z_1^2 Z_3^2Z_2^2 
+ \bar q^2 Z_3^2 Z_1 Z_2 Z_1 Z_2  + \bar q^4 Z_1 Z_2Z_1 Z_3 Z_2 Z_3
\\ \nonumber
& & \bar q^6 Z_3^2Z_2 Z_1Z_2Z_1 + \bar q^4 Z_1 Z_2^2 Z_1 Z_3^2 + 
\bar q^6 Z_2^2 Z_1 Z_3 Z_1 Z_3 + \frac{1}{2} \bar q^2 Z_1Z_2Z_3 Z_1Z_2Z_3 
\\ \nonumber
& & \left. \bar q^4 Z_2 Z_1 Z_2 Z_3 Z_1 Z_3  + \bar q^4 Z_1 Z_3 Z_1 Z_2 Z_3  +
\bar q^2 Z_2 ^2 Z_3 Z_1 Z_3 Z_1 
+ \frac{1}{2} \bar q^6 Z_1 Z_3 Z_2 Z_1 Z_3 Z_2 
\right). 
\end{eqnarray}

Consider the three point function 
\begin{equation} \label{tree3}
 \langle { O}_{(k, k, k)}  \bar {O}_{(k_1, k_1, k_1) }  \bar {O}_{(k_2, k_2, k_2) } 
\rangle 
\end{equation}
where $k= k_1+k_2$. 
We have shown that this three point function 
 does not get renormalized at one loop.  
There is another important property of this three point function. 
The tree level 3 point function  is independent of $\beta$ in the planar limit. 
This can seen by the fact that in the planar limit there is a non-zero 
contribution from the Wick contraction only if the letters in 
${ O}_{(k, k, k)} $ are aligned with the other two. 
Let us focus on one particular term 
$T_k$ in operator ${ O}_{(k, k, k)}$ when the Wick contraction 
is non-zero with the first terms ${\rm Tr} ( \bar Z_1^{k_1} \bar Z_2^{k_1} \bar Z_3^{k_1}), 
{\rm Tr} ( \bar Z_1^{k_2} \bar  Z_2^{k_2} \bar Z_3^{k_2})$ of operators 
$\bar {O}_{(k_1, k_1, k_1) } ,  \bar {O}_{(k_2, k_2, k_2) } $  respectively. 
We call these terms $T_{k_1}$ and  $T_{k_2}$ . 
Now let $S_k$, $S_{k_1}$, $S_{k_2}$ be another set of terms in each 
of the operators which 
have non-zero overlap.  
The number of permutations to obtain $S_k$ from $T_k$ 
   must be equal to the total number of permutations
 to obtain $S_{k_1}$ and $S_{k_2}$ from $T_{k_1}$ and $T_{k_2}$ respectively. 
 From the construction of operators given in (\ref{defcp}), this implies that the 
 relative phase between the contribution of the $T$'s to the three point function
 and the contribution of the $S$'s to the three point function vanishes. 
 Note that the phases in $ \bar {O}_{(k_1, k_1, k_1) },  \bar {O}_{(k_2, k_2, k_2) } $ 
 involve $q$ while the phases in ${ O}_{(k, k, k)}$ involve $\bar q$. 
 Thus other than an over all phase there are no relative phases between any 
 terms in the contractions which give rise to the tree level three point function 
given in (\ref{tree3}).  This over all phase can be removed by an appropriate 
redefinition of say the first operator. 
One can explicitly check this for the the case of $k=2, k_1=k_2 =1$ with the 
operators given in (\ref{exam}).  
Therefore we conclude that the tree level structure constant in the planar limit
is independent of the deformation $\beta$. 

\vspace{.5cm}
\noindent
We summarize the  properties of the three point function given in 
(\ref{tree3}). 
\begin{enumerate}
\item The structure constant does not receive corrections at one loop in the 
t' Hooft limit. 
\item The tree level planar structure constant is independent of the 
deformation $\beta$. 
\item Properties (1) and (2) imply that the structure constant (\ref{tree3}) is 
equal to that in ${\cal N}=4$ Yang-Mills to one loop in t 'Hooft coupling.  
\end{enumerate}
If these properties were to hold to all orders in t'Hooft coupling 
we should expect to see these properties in the Lunin-Maldacena geometry
which is the strong coupling dual of the beta deformed theory.

\section{Structure constants from the Lunin-Maldacena background}

Motivated by the arguments in the previous section we turn towards evaluating structure
constants in the Lunin-Maldacena background which is the strong coupling dual. 
We will use the method developed by  \cite{Zarembo:2010rr}  for evaluating 
the three point function. 
There has been holographic computations of three point functions of  a chiral primary 
with operators dual  to large semi-classical strings in the Lunin-Maldacena 
background  by 
\cite{Ahn:2011dq,Alizadeh:2011yt}. 
The chiral primary involved was restricted  to to be dual to the dilaton in these works
\footnote{ 
Recently in 
\cite{Bozhilov:2013bya}
 the chiral primary involved has been chosen to be
${\rm Tr }(Z^k)$ and ${\rm Tr} ( F^2 Z^k)$. However the vertex operators
chosen for the semi-classical calculations was the same as that of $AdS_5\times S^5$. 
The fat semi-classical string obeyed twisted boundary conditions. }.
Our primary focus is on evaluating the structure constants of the three chiral primary 
given in (\ref{tree3}) and showing that the result is independent 
of the deformation and it  reduces to the same 
evaluated for the holographic dual of ${\cal N}=4$ Yang-Mills. 
We then evaluate the structure constants of the R-current with semi-classical 
string states and verify a Ward-identity. 
 We also evaluate the  structure constants of the 
chiral primary with charge $(k, k, k)$ with other semi-classical string states.

The first step is to evaluate the supergravity mode dual to the chiral primary $O_{(k, k, k)}$. 
This is done in section 4.1. Then  in section 4.2 we  use this to evaluate the three point function 
given in (\ref{tree3})  using the method of  \cite{Zarembo:2010rr}. 
We show that the answer is identical to that of the ${\cal N}=4$ result.
This result together with the one loop results of the previous section suggests that the 
the three point function in (\ref{tree3}) is protected by a
 non-renormalization theorem.

\subsection{The $(k, k, k)$ chiral primary at strong coupling}

In this section we will write down the gravity fluctuations which is dual to 
the chiral primary state ${ O}_{(k, k, k)}$ in the Lunin-Maldacena background.  
Note that the Lunin-Maldacena background is obtained by 
performing a { TsT} transformation on the $AdS_5\times S^5$ geometry 
\cite{Lunin:2005jy,Frolov:2005dj}. 
The strategy we adopt is the following: 
\begin{enumerate}
 \item Identify the chiral primary fluctuations corresponding to the operator 
with $U(1)$ charge ${(k, k, k)}$ in the $AdS_5\times S^5$ background.  
\item  Show that the background together with the  fluctuation 
preserves an $U(1)$ isometry  so that the { TsT}  transformation can be 
performed.
\item Perform the { TsT} transformation on the background  with the 
fluctuations and obtain the Lunin-Maldacena background along with the 
fluctuations.  This yields the super gravity mode dual to the operator
${ O}_{(k, k, k)}$. 
\item As a consistency check we verify that the equations of 
motion of the fluctuations in the Lunin-Maldacena 
background  reduces to that of Klein-Gordan field 
with (mass)${}^2 = k(k-4)$. 
\end{enumerate}

Before proceeding  let us first review how the Lunin-Maldacena background 
is obtained by performing a  { TsT} transformation on the $AdS_5\times S^5$ background. 
We follow \cite{Frolov:2005dj}. 
The $AdS_5\times S^5$ background is given by
\begin{equation}
 ds^2 = \frac{1}{z^2}(  dx^2 + dz^2)  + d\Omega_5^2, \\ \nonumber
\end{equation}
where $d\Omega_5^2$ is the metric on the sphere.
 Let us  parametrize the 
sphere as 
\begin{eqnarray}
& & d\Omega_5^2=\sum_{i}^3 dr_i^2+\sum_{i}^3 r_i^2 d \tilde {\tilde{\phi}}_i^2, \\ \nonumber
& & r_1^2 + r_2^2 + r_3^2 = 1, \\ \nonumber
& & r_1 = \cos\alpha \equiv c_\alpha, 
\qquad r_2 = \sin\alpha\cos\theta \equiv s_\alpha c_\theta, \qquad r_3 = \sin\alpha\sin\theta
\equiv s_\alpha s_\theta. 
\end{eqnarray}
 There is also the background $RR$ 4-from 
which is given by 
\be
\label{rr4s}
C_4=4(w_4+w_1\we d\tilde\tp_1\we d\tilde\tp_2\we d\tilde\tp_3),
\ee
where
\begin{eqnarray}
& & dw_4=\w_{AdS_5} ,\qquad dw_1=\sqrt{g_{s^5}}d\a\we d\theta=c_\a s^3_\a s_\theta c_\theta d\a\we d\theta, \\  \nonumber
& & w_1=\frac{1}{4}s^4_\a s_\theta c_\theta d\theta.
\end{eqnarray}
The dilaton background is constant and we have set $e^{\Phi}=1$ and the radius of $AdS$ to be unity.
We will re-introduce the 
constant mode of the dilaton  and the radius of $AdS$ in the sigma model coupling as in 
\cite{Zarembo:2010rr}. 

\vspace{.5cm}
\noindent
The TsT dual is obtained as follows \cite{Frolov:2005dj}:
\begin{enumerate}
\item
 First perform the following  co-ordinate transformation 
\be
\label{co-trans}
\tilde {\tp}_1=\tilde \tvp_3-\tilde\tvp_2, \qquad \tilde {\tp}_2=\tilde \tvp_3+\tilde\tvp_2+\tilde\tvp_1,\qquad \tilde{\tp}_2=\tilde \tvp_3-\tilde\tvp_1.
\ee
Substituting these transformations in the metric of the $S^5$ we see that the metric is no longer 
diagonal. The RR 4-form also is written in these co-ordinates. 
\item
We then perform T-duality along the $\tilde\tvp_1$ direction. 
This is done by applying the standard T-duality rules given in \cite{Meessen:1998qm}. 
We are now in the  type IIA theory. 
Let us now label the $3$ circles as $\tvp_1, \tvp_2$ and $\tvp_3$. 
\item
Then the following shift is done by replacing
\begin{equation}
\tvp_2 \rightarrow \tvp_2 + \gamma\tvp_1.
\end{equation}
\item
The next step is to again perform a T-duality along the $\tvp_1$ direction. 
Now we are back in type IIB theory,  we label the circles now as $\vp_1, \vp_2, \vp_3$. 
\item
Finally we undo the co-ordinate transformation in (\ref{co-trans}) by performing the inverse
transformation given by 
\be
\varphi_1=\frac{\phi_2+\phi_1-2\phi_3}{3},\qquad \varphi_2=\frac{\phi_2+\phi_3-2\phi_1}{3},\qquad\varphi_3=\frac{\phi_1+\phi_2+\phi_3}{3}.
\ee
\end{enumerate}
The result of this TsT transformation is the following solution of type IIB gravity in the 
string frame
\begin{eqnarray}\label{tsts1}
ds^2 &=& \frac{1}{z^2} \left( dx^2 + dz^2 \right) + \sum_{i=1}^3( dr_i^2 + G r_i^2 d\phi_i^2) 
+ \gamma^2 G r_1^2 r_2^2 r_3^2 \sum_i d\phi_i \sum_j d\phi_j , \\ \nonumber
G&=&\frac{1}{1+\gamma^2(r_1^2r_2^2+r_2^2r_3^2+r_3^2r_1^2)} . \\ \nonumber
\end{eqnarray}
The background also contains the dilaton which is given by 
\begin{equation}\label{tsts2}
\Phi= \frac{1}{2} \log G.
\end{equation}
The anti-symmetric NS-NS 2-form fields which are turned on are given by 
\be\label{tsts3}
b_{\phi_1\phi_2}=G\g r_1^2r_2^2,\quad b_{\phi_2\phi_3}=G\g r_2^2r_3^2, \quad b_{\phi_3\phi_1}=G\g r_3^2r_1^2.
\ee
The component of the RR 2-form fields which are turned on are given by  following 
transformation rules 
\begin{eqnarray}\label{rrtrans}
 C^{2\g}_{\a\theta}&=&\g \left( -C^4_{\a\theta {\tilde \tp_2}{\tilde\tp_3}}+C^4_{\a\theta {\tilde \tp_1}{\tilde\tp_3}}-C^4_{\a\theta {\tilde \tp_1}{\tilde\tp_2}}\right),\\
C^{2\g}_{\theta \phi_i}&=&-\g C^4_{\theta {\tilde \tp_1 }{\tilde \tp_2}{\tilde \tp_3 }}, \\
C^{2\g}_{\a \phi_i}&=&-\g C^4_{\a {\tilde \tp_1 }{\tilde \tp_2}{\tilde \tp_3 }}. \\ \nonumber
\end{eqnarray}
These transformation rules can be obtained by a straight forward application of T-duality 
rules given in \cite{Meessen:1998qm}.  
Here the values for the components  on the right hand 
of the above equations can be read out from the expression for the 
4-form given in (\ref{rr4s}). 
Implementing the RR transformation rules we find the following RR 2-form components  turned on
\begin{equation} \label{tsts4}
C^{2\gamma}_{\theta\phi_i} = -\g \sin^4\alpha \sin\theta \cos\theta.
\end{equation}
Finally applying the TsT rules we find that  the anti-symmetric 4-form RR field is given by 
\begin{equation}\label{tsts5}
C = 4( \omega_4 +  G w_1 d\phi_1\wedge d\phi_2 \wedge d \phi_3 ),
\end{equation}
where 
\begin{eqnarray}
d\omega_4 = \omega_{AdS_5} , \qquad  \omega_1 = \frac{1}{4} \sin^4\alpha \sin\theta\cos\theta d\theta
\\ \nonumber
 d\omega_1 = \cos\alpha\sin^3\alpha\sin\theta\cos\theta  = \sqrt{g_S^5}  d\alpha \wedge d\theta.
 \end{eqnarray}
 Note that  the fields of background after the TsT transformation 
 are all  parametrized by the deformation parameter $\gamma$. 
 This background satisfies the type IIB equations of motion which can be obtained 
 from the action 
 \begin{eqnarray} \nonumber
S & =& \frac{1}{2\k_{10}^2}\int d^{10} x\sqrt{-g_E} \left( R- \frac{1}{2}\partial_M \Phi \partial^M \Phi -\frac{e^{-\Phi}H_3^2}{2.3!}- \frac{1}{2}e^{2\Phi}F_1^2 - \frac{e^{\Phi}F_3^2}{2.3!}-\frac{1}{4.5!}F_5^2 \right) \\ 
&&  -\frac{1}{2 \k _{10}^2}\int C_4 \wedge H_3\wedge F_3.
\end{eqnarray} 
Note that this action is written in Einstein frame and 
$2\k_{10}^2= (2\pi)^7 \alpha^{\prime 4}$. 
The relation between the string frame and the Einstein frame is given by 
\be \label{reles}
g_{MN}^E=e^{-\Phi/2}g_{MN}^S.
\ee
The equations of motion is further supplemented
by the self duality constraint on the $5$-form field strength. 
 When $\gamma=0$, the solution given in (\ref{tsts1}), (\ref{tsts2}), 
 (\ref{tsts3}), (\ref{tsts4}) and (\ref{tsts5})  reduces to the usual $AdS_5\times S^5$. 
 The deformation 
 $\gamma$ is identified with $\beta/2\pi$ of the beta deformed theory 
 \cite{Lunin:2005jy}. 
 
 Let us now examine the fluctuations which correspond to the chiral primaries in the 
 $AdS_5\times S^5$ background \cite{Kim:1985ez,Lee:1998bxa}.  
 We perturb the metric and the RR 4-form by 
 \begin{equation}
 g_{MN} = g_{MN}^{(0)} + \varepsilon h_{MN}, \qquad \tilde C_4  = C_4^{(0)}  + \varepsilon a_4, 
 \end{equation}
 where the non-zero components of the  perturbations  $h_{MN}$ and $a_4$ are defined by 
 \cite{Lee:1998bxa}\footnote{See \cite{Zarembo:2010rr} for a brief summary.}
 \begin{eqnarray}
\label{s5fluc}
 h_{mn}&=&\frac{2}{{\cal N}_k(k+1)} Y_I[2\nabla_m\nabla_n-k(k-1)g_{mn}]s_I(x, z)  ,\\
h_{\a\b}&=&\frac{2 k g_{\a \b}Y_I s_I (x, z) }{{\cal N}_k}, \\
a_{\a\b\g\d}&=& -4 \sqrt{g_{s^5}} \frac{\e_{\a\b\g\d\e}\nabla^\e Y_I s_I(x, z) }{{\cal N}_k}, \\
a_{mnpr}&=& 4 \sqrt{g_{AdS_5}} \frac{\e_{mnprs} Y_I \nabla^s s_I(x, z) }{{\cal N}_k} .\\
\end{eqnarray}
The labels $m, n, \cdots $ take values from $0, 1, \cdots 4$ which label the $AdS_5$ directions. 
The labels $ \alpha, \beta, \cdots$ take values from $5, 6, \cdots 9 $ which label the 
$S^5$ directions. $Y_I$ are spherical harmonics on the $S^5$, 
I labels the various spherical harmonics. The spherical 
harmonics satisfy the equation
\begin{equation}
\nabla_{\alpha}\nabla^\alpha Y_I  + k( k+4) Y_I = 0
\end{equation}
Thus $k$ refers to the rank of the spherical harmonic.
For future reference we  write down the non-zero components of the 
RR 4-form on $S^5$  along with the fluctuations explicitly:
\begin{eqnarray}\label{rr4p}
C^4_{\theta {\tilde \tp_1}{\tilde \tp_2}{\tilde \tp_3}}&=&s^4_\a s_\theta c_\theta -4\varepsilon \frac{\sqrt{g_{S^5}}}{{\cal N}_k}s_I\partial_{\a}Y_I ,\\ \nonumber
C^4_{\a \ttp_1\ttp_2\ttp_3}&=& 4\varepsilon \frac{\sqrt{g_{S^5}}}{{\cal N}_k} s_I \partial_{\theta}Y_I g^{\theta \theta} ,\\ \nonumber 
C^4_{\a\theta {\tilde \tp_2}{\tilde \tp_3}}&=& -4\varepsilon \frac{\sqrt{g_{S^5}}}{{\cal N}_k} s_I \partial_{\tilde \tp_1}Y_I g^{\tilde \tp_1 \tilde \tp_1} ,\\ \nonumber
C^4_{\a \theta \ttp_1\ttp_3} &=& 4\varepsilon \frac{\sqrt{g_{S^5}}}{{\cal N}_k} s_I \partial_{\tilde \tp_2}Y_I g^{\ttp_2 \ttp_2} ,\\ \nonumber
C^4_{\a \theta \ttp_1\ttp_2} &=& -4\varepsilon \frac{\sqrt{g_{S^5}}}{{\cal N}_k} s_I \partial_{\tilde \tp_3}Y_I g^{\ttp_3 \ttp_3} ,
\end{eqnarray}
with $g^{\a\a}=1, g^{\th \th}=\operatorname{cosec}^2{\a}, g^{\tilde \tp_i \tilde \tp_i}=r_i^{-2}$ $(i=1,2,3)$.
 A simple example of a spherical harmonic on $S^5$  is given by 
\begin{equation} \label{sphehar}
Y_1 = \left( \frac{r_1}{ \sqrt{2} } \right) ^k \exp ( ik\tilde\tp_1) .
\end{equation}
This spherical harmonic carries a charge $k$ along the direction $\tilde\tp_1$. 
Our conventions for
the $\epsilon$ tensor are $\epsilon_{01234}= \epsilon_{56789}=1$. 
Now expanding the type IIB equations of motion to the linear order in 
fluctuation about the $AdS_5\times S^5$ background it can be seen that the 
amplitude of the fluctuations $s_I$ satisfy the 
minimally coupled massive scalar field in  $AdS_5$ which is given by 
\begin{equation}\label{minscal}
\nabla_m\nabla^m s_I - k(k-4) s_I =0.
\end{equation}
The normalization ${\cal N}_k$ is 
determined by holographically evaluating the two point function of the dual operator to this fluctuation
and demanding it to be unity. For the spherical harmonic given in (\ref{sphehar}) this results in 
the following \cite{Zarembo:2010rr}. 
\begin{equation}
({\cal N}_k^{(1)})^2  = \frac{N^2 k(k-1)}{2^{k-3} \pi^2 ( k+1)^2}.
\end{equation}
The superscript here refers to the fact that the normalization corresponds to 
the harmonic in (\ref{sphehar}). 
Note that the fluctuations satisfy the minimally coupled scalar 
equation in (\ref{minscal}) with  (mass)${}^2=k(k-4)$ .The $U(1)$ charge carried
by the spherical harmonic in (\ref{sphehar}) is given in $(k, 0, 0)$. 
These two facts allow us to  conclude that this supergravity mode  is dual to the 
chiral primary ${\rm Tr} ( Z_1^k)$. 

We now  examine a chiral primary fluctuation in the $AdS_5\times S^5$ background
which preserves the $U(1)$ isometry along $\tilde \varphi_1$ on which the $T$ duality 
is done. 
Note that the spherical harmonic given in (\ref{sphehar}) does not preserve 
the $U(1)$ isometry.  The reason is that 
 after the  shift $\tilde \varphi_2 \rightarrow  + \gamma\tilde \varphi_1$, 
the fluctuation depends on the direction $\tilde \varphi_1$ and thus breaks the
$U(1)$ isometry.  The following spherical harmonic of rank $k$ preserves the $U(1)$ isometry.
\begin{equation} \label{harm2}
Y_2  =\frac{1}{2^{\frac{k}{2}} } ( r_1 r_2 r_3)^{\frac{k}{3} } \exp\left(   \frac{i k}{3} 
( \tilde\tp_1 + \tilde\tp_2   + \tilde\tp_3) \right).
\end{equation}
Here $k$ is a multiple of $3$ so that the  the function is periodic in the angles.  
Using the co-ordinate transformation in 
(\ref{co-trans}) we see that this harmonic just depends on the angle $\tilde\varphi_3$. 
Thus it preserves the $U(1)$ isometry along which the TsT transformation is done. 
From the $U(1)$ quantum numbers of this harmonic we see that it is dual to the
operator ${ O }_{\frac{k}{3} \frac{k}{3} \frac{k}{3} }$. 
Before proceeding we will determine normalization constant ${\cal N}_k^{(2)}$
corresponding to the harmonic in (\ref{harm2}). 
The normalization is determined 
by requiring that the holographic two point function be normalized to 
unity. From  the analysis of \cite{Lee:1998bxa} we see that  the normalization satisfies the following 
property
\be
\label{N1}
({\cal N}_k^{(I)} )^{2} =f(k) z_I(k),
\ee
where $f(k)$ is quantity independent of the choice of spherical harmonics. But $z_I(k)$ is given by 
\cite{Lee:1998bxa}
\be
\int Y_{I_1}Y_{I_2}=z(k) \d^{I_1 I_2}.
\ee
For the spherical harmonic given in (\ref{sphehar}) 
it can be seen that
\be
z_1(k)=\frac{2^{-k}8\pi^3}{4(2+k)(k+1)}.
\ee
For the second choice of spherical harmonic given in (\ref{harm2})
we obtain 
\be
z_2(k)=\frac{8 \pi^3 2^{-k/2} (\G(1+k/3) ) ^3}{4 \G(k+3)}.
\ee
Then from (\ref{N1}) one can write
\begin{eqnarray} \label{N2}
( {\cal N}_k^{(2)})^2 &=&  ( {\cal N}_k^{(1)})^2 \frac{z_2(k)}{z_1(k)}=
( {\cal N}_k^{(1)})^2 \frac{( \G(1+k/3) ) ^3}{\G(k+1)}.
\end{eqnarray}

We now consider the $AdS_5\times S^5$ background along with the fluctuations given in 
(\ref{s5fluc}) with the harmonic (\ref{harm2}). 
Since the full background preserves the $U(1)$ isometry  we can
perform the TsT transformation. The TsT transformation maps solutions of 
equations of motion to solutions.  Thus the fluctuations obtained after the TsT transformation 
will satisfy the linearized equations of motion about the Lunin-Maldacena background 
if (\ref{minscal}) is true. 
We now write down the fluctuations about the Lunin-Maldacena background 
obtained by performing  the TsT transformation of the chiral primary corresponding 
to the harmonic in  (\ref{harm2}).
The background metric and the fluctuations are given by 
\begin{eqnarray} \label{tstf1}
G_{mn}  &=& g_{mn} + \epsilon \frac{2}{{\cal N}_k(k+1)} Y_2[2\nabla_m\nabla_n-k(k-1)g_{mn}]s_2,
\\ \nonumber
G_{\a\a}&=&1+\varepsilon\frac{2 k Y_2s_2}{{\cal N}_k}  \\ \nonumber
G_{\theta \theta}&=& \sin^2 \a \left(1+\varepsilon\frac{2 k Y_2s_2}{{\cal N}_k}\right) ,\\ \nonumber
G_{\phi_1\phi_1}&=& G(r_1^2+\g^2r_1^2r_2^2r_3^2) \\ \nonumber
& & +\varepsilon\frac{2 k Y_2s_2}{{\cal N}_k} 
\left(\g^4( r_1^4 r_2^4 r_3^2+  r_1^4 r_2^2 r_3^4+  r_1^2 r_2^4 r_3^4)+ 
\g^2 (2r_1^2 r_2^2 r_3^2-  r_1^4 r_2^2-  r_1^4 r_3^2)+ r_1^2\right)G^2 ,\\ \nonumber
G_{\phi_2\phi_2}&=& G(r_2^2+\g^2r_1^2r_2^2r_3^2) ,\\ \nonumber
& & +\varepsilon\frac{2 k Y_2s_2}{{\cal N}_k}\left( \g^4 (r_1^4 r_2^4 r_3^2+  r_1^4 r_2^2 r_3^4+ r_1^2 r_2^4 r_3^4)+  \g^2 (2r_1^2 r_2^2 r_3^2-  r_2^4 r_1^2-  r_2^4 r_3^2)+ r_2^2\right)G^2,\\ \nonumber
G_{\phi_3\phi_3}&=&G(r_3^2+\g^2r_1^2r_2^2r_3^2) ,\\ \nonumber
& & +\varepsilon\frac{2 k Y_2s_2}{{\cal N}_k} \left( \g^4 (r_1^4 r_2^4 r_3^2+  r_1^4 r_2^2 r_3^4+ r_1^2 r_2^4 r_3^4)+  \g^2 (2r_1^2 r_2^2 r_3^2-  r_3^4 r_1^2-  r_3^4 r_2^2)+ r_3^2\right)G^2,\\ \nonumber
G_{\phi_1 \phi_2}&=&G\g^2r_1^2r_2^2r_3^2 ,\\ \nonumber
& & +\varepsilon\frac{2 k Y_2s_2}{{\cal N}_k} \left( \g^4( r_1^4 r_2^4 r_3^2+ r_1^4 r_2^2 r_3^4+  r_1^2 r_2^4 r_3^4)+3  \g^2 r_1^2 r_2^2 r_3^2\right)G^2, \\  \nonumber
G_{\phi_2 \phi_3}&=&G_{\phi_1 \phi_3}=G_{\phi_1 \phi_2}.
 \end{eqnarray}
 In the above  equations and the rest of the paper it is understood that the normalization 
 ${\cal N}_k$ refers to  ${\cal N}_k^{(2)}$  which is given in (\ref{N2}). 
 The background NS B-field along with the fluctuations are given by 
 \begin{eqnarray}\label{tstf2}
 B_{\phi_1 \phi_2}&=& G\g r_1^2 r_2^2 +2\varepsilon\frac{2 k Y_2s_2}{{\cal N}_k} G^2 \g r_1^2 r_2^2 ,\\ \nonumber
B_{\phi_2 \phi_3}&=& G\g r_2^2 r_3^2 +2\varepsilon\frac{2 k Y_2s_2}{{\cal N}_k} G^2 \g r_2^2 r_3^2, \\ \nonumber
B_{\phi_3 \phi_1}&=& G\g r_3^2 r_1^2 +2\varepsilon\frac{2 k Y_2s_2}{{\cal N}_k} G^2 \g r_3^2 r_1^2 .\\ 
 \end{eqnarray}
 The background dilaton with the fluctuation is given by 
 \begin{equation}\label{tstf3}
\Phi=\frac{{\rm Log}G}{2}+\varepsilon \frac{2kY_2 s_2}{{\cal N}_k} (G-1) .
\end{equation}
The RR 2-form components  with the fluctuations 
are   given by the relations  given in 
(\ref{rrtrans}) on the RR 4-components  before the TsT transformation given in (\ref{rr4p}). 
One can also obtain the RR 4-form along with the fluctuation after the TsT transformation.
However as will be seen in the next section 
our analysis requires only the NS fields and their fluctuations. 

As a simple check of the fact that we have got the right set of fluctuations note that 
for $\gamma =0$, the background as well as the fluctuations reduces to the situation in 
$AdS_5\times S^5$. 
As a  further non-trivial consistency check  we verify that  the amplitude of the fluctuation $s_2$ 
satisfies the minimally coupled scalar equation in $AdS_5$  with (mass)${}^2 = k (k-4)$. 
To do this we first examine the dilaton equation of type IIB. In the Einstein frame  this is 
given by 
\be\label{eomd}
\frac{1}{\sqrt{-g_E}}\partial_M (g^{MN}_E\sqrt{-g_E} \partial_N \Phi)=e^{2\Phi}F_1^2+\frac{1}{2.3!}(e^\Phi F_3^2-e^{-\Phi}H_3^2).
\ee
We then convert our background and the fluctuations 
given in (\ref{tstf1}), (\ref{tstf2}), {\ref{tstf3} 
to the Einstein frame 
using (\ref{reles}).    This is substituted in the equation of motion for the 
dilaton (\ref{eomd}).  We then 
expand the 
LHS of the equation to the linear order in $\varepsilon$. After a tedious manipulation
using Mathematica it can indeed be shown that if the amplitude $s_2$ satisfies 
(\ref{minscal}) then the type IIB  dilaton equation  is satisfied to the linear order in 
the fluctuations.

\subsection{3-pt functions of chiral primaries at strong coupling}

Having constructed the super-gravity mode   dual to the chiral primary 
 operator ${\cal O}_{kkk}$ we are in a position
to evaluate the structure constants  of three chiral primaries defined in 
(\ref{tree3}). For this purpose we will use the method developed by 
\cite{Zarembo:2010rr}  which  evaluates the structure constant of 
a super-gravity mode with two semi-classical strings states in terms of 
a world-sheet amplitude. 
Let the metric and the 
NS-B field fluctuations of the super-gravity mode in the string frame be given by
$h_{MN}^I$ and $b_{MN}^I$, where $I$ labels 
the mode. Then the structure constant is extracted from the amplitude
\cite{Zarembo:2010rr}. 
\begin{equation}\label{amp1}
\tilde C_{I}  = -\frac{\sqrt{2 ( \Delta_I -1) \lambda}}{8\pi^2} 
\int d^2 \sigma \sqrt{\tilde h} \partial_a X^M \partial_b X^N\left( \tilde h^{ab}  h_{MN}^I + i \epsilon^{ab}
b_{MN}^I  \right) z^{\Delta_I}.
\end{equation}
Here $\tilde h_{ab}$ is the world sheet metric which in the conformal gauge can be chosen to be
$\delta_{ab}$. 
$h_{MN}$ and $b_{MN}$ are the fluctuations of super-gravity mode stripped without the 
amplitude $s_I$. 
$X^M(\sigma^1, \sigma^2)$ are classical solutions to the world sheet sigma model 
in the back-ground of interest. They must have the property that they originate from the boundary
of $AdS$ and end at another point on the boundary. The dependence on the 
 distance which separates the two points 
has to be extracted from the amplitude to read off the structure constant.  
$\Delta_I$ is the conformal dimension of the operator dual to the super-gravity mode. 
$z$ is the radial co-ordinate of $AdS_5$ and $\lambda$ is the t'Hooft coupling. 
Note that here we have introduced the constant value of the dilaton and the $AdS$ radius measured
in units of the string length as the t'Hooft coupling which serves as the string tension. 
The classical solutions are in general complex as they are analytical continuation of 
solutions in Minkowski world sheet.

From the expression for the structure constant at strong coupling 
in (\ref{amp1}) we see that there are three
ingredients to evaluate them. 
\begin{enumerate}
\item The background solution in the string frame represented by the metric  $g_{MN}$ and the 
NS B-field $b_{MN}$.  
\item The fluctuations  of the metric and the NS B-field $h_{MN}, \beta_{MN}$ which are 
dual to the chiral primary operator we are interested. 
\item The semi-classical sigma model solution dual to the large operators represented by the 
world sheet solutions $X^{M}(\sigma^1, \sigma^2 )$ in the background we are interested in. 
\end{enumerate}
The background of interest in this paper is the Lunin-Maldacena solution which is the holographic 
dual to the beta deformed theory. 
We will now proceed to apply  the expression (\ref{amp1})  to show that the structure constants
of the three chiral primaries defined in (\ref{tree3}) in 
the beta deformed theory at strong coupling  is independent of the deformation 
$\gamma$ and is identical to that of ${\cal N}=4$ Yang-Mills. 
This demonstrates that these structure constants  in the beta deformed theory 
at strong coupling are identical to that
evaluated at the tree level, thus providing evidence for a non-Renormalization theorem. 

Substituting the fluctuations of the metric and the NS-B field into (\ref{amp1}) we obtain
\be
\label{struc1}
\begin{split}
{\cal C}=&\sqrt{\frac{\G(k+1)}{\G(1+k/3)^3}}\frac{2^{\frac{k}{2}-3} \sqrt{k \l}(k+1)}{\p N} \times \\ 
&\int d^2 \s Y_2 [z^{k-2}(\partial x)^2-  z^{k-2}(\partial z)^2- z^k(H_{\a\b}\partial_aX^\a \partial ^a X^\b )\\
&-i2z^k(\beta_{12}\partial_a \phi^1\partial_b \phi^2+\beta_{13}\partial_a \phi^1\partial_r \phi^3+\beta_{23}\partial_a \phi^2\partial_b \phi^3)\e^{ab}],
\end{split}
\ee
where we have used $\nabla_x\nabla_x z^k=-kz^{k-2}$ and $\nabla_z\nabla_z z^k=k^2z^{k-2}$. 
Note that the world sheet is Euclidean and 
\begin{eqnarray}\label{amflu}
H_{\alpha\alpha} &=& 1, \\ \nonumber
H_{\theta\theta} &=& \sin^2\alpha, \\ \nonumber
H_{\phi_1\phi_1} &=&   \left(\g^4( r_1^4 r_2^4 r_3^2+  r_1^4 r_2^2 r_3^4+  r_1^2 r_2^4 r_3^4)+ 
\g^2 (2r_1^2 r_2^2 r_3^2-  r_1^4 r_2^2-  r_1^4 r_3^2)+ r_1^2\right)G^2, \\ \nonumber
H_{\phi_2\phi_2}&=& 
\left( \g^4 (r_1^4 r_2^4 r_3^2+  r_1^4 r_2^2 r_3^4+ r_1^2 r_2^4 r_3^4)+  \g^2 (2r_1^2 r_2^2 r_3^2-  r_2^4 r_1^2-  r_2^4 r_3^2)+ r_2^2\right)G^2,\\ \nonumber
H_{\phi_3\phi_3}&=&
 \left( \g^4 (r_1^4 r_2^4 r_3^2+  r_1^4 r_2^2 r_3^4+ r_1^2 r_2^4 r_3^4)+  \g^2 (2r_1^2 r_2^2 r_3^2-  r_3^4 r_1^2-  r_3^4 r_2^2)+ r_3^2\right)G^2,\\ \nonumber
H_{\phi_1 \phi_2}&=&
 \left( \g^4( r_1^4 r_2^4 r_3^2+ r_1^4 r_2^2 r_3^4+  r_1^2 r_2^4 r_3^4)+3  \g^2 r_1^2 r_2^2 r_3^2\right)G^2, \\  \nonumber
H_{\phi_2 \phi_3}&=&H_{\phi_1 \phi_3}=H_{\phi_1 \phi_2}, \\ \nonumber
\beta_{12}  &=&  G^2 \g r_1^2 r_2^2 ,\\ \nonumber
\beta_{23} &=& G^2 \g r_2^2 r_3^2, \\ \nonumber
\beta_{31} &=& G^2 \g r_3^2 r_1^2 .
\end{eqnarray}

Let us now determine the semi-classical solution dual to the chiral 
primary of interest. These are geodesics which have equal angular momentum along the 
three $U(1)$'s.  We first write down these classical solutions. 
The trajectory in the the Lunin-Maldacena geometry is given by 
\begin{eqnarray}\label{clas1}
 x&=&R {\rm tanh}\k \t,\quad z=\frac{R}{{\rm cosh}\k\t} , \qquad R=\frac{|x_1-x_2|}{2} \\ \nonumber
\phi_1&=&\phi_2=\phi_3=i \w \t, \qquad r_1=r_2=r_3=\frac{1}{\sqrt 3}.
\end{eqnarray}
Note that only when $r_i  = \frac{1}{\sqrt{3}}$ the
equations of motion are satisfied. 

The Virasoro constraint for this solution results in the following equality 
\be
\k^2=\w^2.
\ee
We can use Noether's prescription to obtain the three $U(1)$ charges for this solution  
which is given by  
\be
J_1=J_2=J_3=\sqrt{\l} \frac{\w}{3}.
\ee
The energy for the solution is given by 
\be
E=\sqrt{\l} \k,
\ee
Using the Virasoro constraint we see that the solution has the property 
$E^2 = J^2$, where $J$ is the sum of the three $U(1)$ charges.  Therefore this geodesic is 
the semi-classical solution dual to the chiral primary $O_{\frac{J}{3}, \frac{J}{3}, \frac{J}{3}}$. 

We now substitute the classical solution (\ref{clas1}) into the integral given in  (\ref{struc1}). 
The following simplification can be observed
\be
\label{bmn3}
\frac{(\partial x)^2-(\partial z)^2}{z^2}=\k^2(-1+2 \operatorname{sech} ^2\k \t).
\ee
Furthermore on substituting the classical solution (\ref{clas1}) into the part of the integral 
which depends on the deformed sphere we obtain the following non-trivial simplification
\be
\label{bmn4}
H_{\a\b}\partial_aX^\a \partial _b X^\b h^{ab}=-\w^2=-\k^2.
\ee
What this implies is that the dependence of the deformation $\gamma$ completely drops out.
Since the classical solution (\ref{clas1}) is a geodesic the fluctuations of the 
anti-symmetric NS B-fields don't play a role in the integral (\ref{struc1}). 
Using these simplifications and extracting the dependence of the distance 
between the end points of the geodesic in (\ref{struc1})  given by $|x_1-x_2|^k$, 
we obtain the following result for the structure constant
\begin{eqnarray} \label{result}
C^J_{J k}&=&\sqrt{\frac{\G(k+1)}{\G(1+k/3)^3}} \frac{1}{3^{k/2} N}2^{-k-1}J(k+1)\sqrt{k}\k \int^{\infty}_{-\infty} d\tau \frac{e^{-k \k \t}}{\cosh ^{k+2} \k \t},  \\ \nonumber
&=&\sqrt{\frac{\G(k+1)}{\G(1+k/3)^3}}\frac{J \sqrt{k}}{3^{k/2} N}.
\end{eqnarray}

The crucial observation of this calculation is that the result for the structure constant
is completely independent of $\gamma$. Thus if one repeats the calculation with 
$\gamma=0$,  the geometry, the fluctuations as well as the classical
solution reduces to that of the in  the ${\cal N}=4$  dual. The    structure then reduces
to that in ${\cal N}=4$ Yang-Mills. 
The answer is also independent of the coupling $\lambda$. 
Note also the typical dependence of  three point functions of the three chiral primaries
of ${\cal N}=4$ 
given by $J\sqrt{k}/N$ for $J>>k$ seen in \cite{Lee:1998bxa}. 
Now from the tree level and the one loop calculations of the previous section also 
we deduced that the structure constants of these chiral primaries in the beta deformed
theory is identical to that of ${\cal N}=4$ Yang-Mills. 
The one loop observations together with the same observations of the behaviour 
of the structure constant at strong coupling
 provides evidence that these structure constants are protected by 
  non-renormalization theorem.

\section{Structure constants involving massive  strings}

In this section we evaluate structure constants involving at least $2$ arbitrary semi-classical string 
states. 
We first review how the  Ward-identity satisfied by the R-current determines the 
structure constant involving 
the R-current and two arbitrary scalars 
to all orders in the coupling constant.  In section 4.2  we verify this prediction 
by evaluating this structure constant  in the Lunin-Maldacena geometry. 
For the $AdS_5\times S^5$ geometry this prediction was verified recently in 
\cite{Georgiou:2013ff}. 
 Finally in section 4.3 we evaluate the structure constant involving the $(k, k, k)$ chiral primary and 
operators dual to a rigid rotating string at strong coupling.

\subsection{Structure constant from a Ward identity}

In a conformal field theory, the 3 point function involving a vector $V_\mu$
and two scalars is  completely determined up to a coefficient by conformal invariance. 
Consider a three point function 
a scalar operator ${\cal O}_{\Delta}$ with $U(1)$  R-charge $J$ 
and its conjugate ${\bar {\cal O}}_{\Delta}$ with the conserved R-current  $V_\mu$. 
The three point function is given by 
\begin{eqnarray}\label{vector}
P_{3\mu}&=&\langle V_{\mu}(x)\,\,{\cal O}_{\Delta}(x_1)\,\,{\bar {\cal O}}_{\Delta}(x_2)\rangle
=\frac{C_J(\lambda)}{x_{12}^{2\Delta-2}(x_1-x)^2(x_2-x)^2}E_{\mu}^x(x_1,x_2), 
\nonumber \\
E_{\mu}^x(x_1,x_2)&=&\frac{(x_1-x)_{\mu}}{(x_1-x)^2}-\frac{(x_2-x)_{\mu}}{(x_2-x)^2},
\end{eqnarray}
where $\Delta$ is the conformal dimension of ${\cal O}_{\Delta}$ and $\mu =0, 1, 2, 3$. 
Since ${\cal O}_\D$  has $U(1)$ charge $J$ it undergoes 
following infinitesimal transformation under the R-symmetry
\be
{\cal O}_\D(x)\rightarrow {\cal O}_\D(x)+ \epsilon \delta {\cal O}_\D(x)={\cal O}_\D(x)+ \epsilon J {\cal O}_\D(x).
\ee
From the R-symmetry of the theory it can be shown that the three point function given in 
 (\ref{vector}) obeys the following Ward identity
\begin{eqnarray}\label{Ward} \nonumber
\langle\partial^{\mu}j_{\mu}^R(x) \,\,{\cal O}_{\Delta}(x_1)\,\,{\bar {\cal O}}_{\Delta}(x_2) \rangle&=& \delta^4(x_1-x)\langle \delta{\cal O}_{\Delta}(x_1)\,\, {\bar {\cal O}}_{\Delta}(x_2)\rangle+\delta^4(x_2-x) \langle {\cal O}_{\Delta}(x_1)\,\, \delta{\bar {\cal O}}_{\Delta}(x_2) \rangle \\ 
&=&J\big(\delta^4(x_1-x)-\delta^4(x_2-x)\big)\langle {\cal O}_{\Delta}(x_1)\, \,{\bar {\cal O}}_{\Delta}(x_2)\rangle. \\ \nonumber
\end{eqnarray}
After differentiating \eqref{vector}  we obtain
\begin{eqnarray}\label{diff-vector}
\partial^{\mu}P_{3\mu}=-C_{J}(\lambda) 2 \pi^2\big(\delta^4(x_1-x)-\delta^4(x_2-x)\big)\frac{1}{x_{12}^{2\Delta}}.
\end{eqnarray}
Comparing \eqref{Ward} and \eqref{diff-vector} results in 
\begin{eqnarray}
\label{C123_R}
C_{J}(\lambda)=-\frac{J}{2 \pi^2}.
\end{eqnarray}
This is an all-loop prediction for the structure constant $C_{J}(\lambda)$. 
In the next section we will verify this at strong coupling in the Lunin-Maldacena 
geometry. 
The method proposed by \cite{Zarembo:2010rr} to evaluate the structure constant
in gravity  relies on taking the limit  $x \rightarrow \infty$. 
Before we proceed we evaluate (\ref{vector})   in  the limit $x\rightarrow \infty$. 
Using $\frac{1}{(x_1-x)^2}\rightarrow\frac{1}{x^2}\left(1+\frac{2 x_1 . x}{x^2}\right)$,  we obtain
\be
E_{\mu}^x(x_1,x_2)=\frac{(x_1-x)_{\mu}}{(x_1-x)^2}-\frac{(x_2-x)_{\mu}}{(x_2-x)^2}\rightarrow \frac{(x_1-x_2)^{\nu}}{\vec x^2}\left(\delta_{\mu \nu}-\frac{2 x_{\mu}x_\nu}{{\vec x}^2}\right).
\ee
Then the leading term of (\ref{vector}) in the $x \rightarrow \infty$ limit is given by 
\be \label{vector2}
\langle j_{\mu}^R(x) \,\,{\cal O}_{\Delta}(x_1)\,\,{\bar {\cal O}}_{\Delta}(x_2) \rangle=C_{J}(\lambda) \frac{x_{12}^{\nu}}{x_{12}^{2\Delta-2}} \frac{1}{(x^2)^3}\left(\delta_{\mu \nu}-\frac{2 x_{\mu}x_\nu}{{ x}^2}\right).
\ee

\subsection{Ward identity at strong coupling }

To evaluate the three point function given in (\ref{vector}) at strong coupling we first need 
to obtain the supergravity mode dual to the R-currents. 
To obtain this mode in the Maldacena-Lunin geometry we first review the 
case for $AdS_5\times S^5$ \cite{Kim:1985ez,Georgiou:2013ff}. 
This mode is combination of  fluctuations of the 
metric and the $4$-form potential. 
The metric fluctuation has  one index on the  internal space $S^5$ and one along the 
$AdS_5$ directions. The  4-form potential has one index along the $AdS_5$ directions 
and the remaining indices along $S^5$. 
The 4-form potential couples to the 4-fermion term in the sigma model action and therefore 
its contribution will be suppressed by $O(1/\sqrt{\lambda})$ compared to that from the 
metric fluctuation. This is because   these terms  contribute   only when loops in the sigma model 
coupling is considered. 
Therefore we examine only the metric fluctuation.
Expanding the metric fluctuations in terms of vector harmonics on the $S^5$ we obtain 
\begin{equation}\label{vecf}
h_{m\alpha} = \sum_I B_m(x) Y_\alpha^I(\Omega).
\end{equation}
where $I$ now runs over the vector harmonics which correspond to the 
the $15$ isometries of $S^5$. 
We will focus on the $3$  $U(1)$ isometries which are preserved by the $TsT$ action 
and which correspond to the $3$ R-currents of the beta deformed theory. 
These isometries are rotations which correspond to angular 
shifts  in the $\tilde\tp_1, \tilde\tp_2, \tilde\tp_2$ direction. 
The vector harmonics corresponding to these can be extracted from their 
Killing vectors. These are given by 
\begin{eqnarray}
Y_{\tilde\tp_1} ^1= r_1^2 =  \cos^2 \alpha
\qquad Y_{\tilde\tp_2} ^2= r_2^2 = \sin^2\alpha\cos^2\theta, \qquad
Y_{\tilde\tp_3}^3 = r_3^3= \sin^2 \alpha\sin^2\theta.
\end{eqnarray}
Note that none  of these depend explicitly on the co-ordinates $\tilde\tp_1, \tilde \tp_2, \tilde \tp_3$. 
Since these isometries are preserved under the TsT transformation we perform the TsT transformation 
on the background $AdS_5\times S^5$ metric together with the fluctuations given in
(\ref{vecf}).  As we have discussed earlier,  for evaluating the  structure constant 
to  the leading order in the coupling $\sqrt{\lambda}$ it is sufficient to keep 
track of only the metric fluctuation and the NS-B fields  which result after performing the 
TsT transformation. 
We will first focus on the supergravity mode corresponding to the vector harmonic
$Y_{\tilde\tp_1}^1$.  Performing the TsT transformation on the background metric
together with this mode we obtain the following mode in the Lunin-Maldacena
geometry. 
\be \label{suvec}
h_{m \alpha}=B_m W_\alpha, \qquad \beta_{m \alpha}=B_m Z_\alpha,
\ee
where 
\begin{eqnarray}
W_{\phi_1} &=&(r_1^2+\g^2r_1^2r_2^2r_3^2)G,\quad W_{\phi_2}=W_{\phi_3}=\g^2r_1^2r_2^2r_3^2G,
\\  \nonumber
 Z_{\phi_1} &=&0,\quad Z_{\phi_2}=\g r_1^2r_2^2G, \quad Z_{\phi_3}=-\g r_1^2r_3^2G.
\end{eqnarray}

Now that we have the fluctuation which is dual to  the R-current we can proceed to 
evaluate the  three point function in (\ref{vector}). 
Following the same method developed by \cite{Zarembo:2010rr} and implemented for the case of 
the three point function involving vectors in \cite{Georgiou:2013ff},  we evaluate the amplitude 
\begin{eqnarray}
\label{expB}
\langle B_\mu(\vec x)  \rangle &=&\frac{\langle j_\mu(\vec x) \,\,{\cal O}_{\Delta}(x_1)\,\,{\bar {\cal O}}_{\Delta}(x_2)  \rangle}{\langle{\cal O}_{\Delta}(x_1)\,\,{\bar {\cal O}}_{\Delta}(x_2) \rangle}  \\ \nonumber
&=&\big\langle B_\mu( x,z'=0)\frac{1}{{\cal Z}_{string}}\int DX \,\, e^{-S_{string}}, 
\big\rangle_{bulk}.
\end{eqnarray}
where  $x$ refers to the co-ordinates on the boundary of the $AdS_5$. 
$S_{string}$ is given by 
\begin{equation}
S_{string} = \frac{\sqrt{\lambda}}{4\pi} 
\int d^2 \sigma 
( \partial^a X^M \partial_b X^N h_{MN} + i \epsilon^{ab} \partial_a X^M\partial_b X^N \beta_{MN} 
) 
\end{equation}
and $h_{MN}, \beta_{MN}$ is the supergravity mode corresponding to the R-current given in 
(\ref{suvec}). 
 Then (\ref{expB}) can be written as
\be
\label{expB1}
\begin{split}
\langle B_\mu( x)  \rangle=-\frac{2\sqrt{\lambda}}{4 \pi}\int d^2&\sigma\,\,[(\partial_{\tau}X^{m}\partial_{\tau}X^{\beta}+\partial_{\sigma}X^{m}\partial_{\sigma}X^{\beta})W_{\beta}+  i\e^{ab}\partial_{a}X^{m}\partial_{b}X^{\beta} Z_\b ] \\
&\times \big\langle B_\mu( x,z'=0) B_{m}(y, z)\big\rangle_{bulk}, 
\end{split}
\ee
where 
\begin{equation}
\big\langle B_\mu( x,z'=0) B_{m}(y,  z )\big\rangle_{bulk}  = G_{m\mu}( y, z; x) 
\end{equation}
is the bulk to boundary propagator 
for vectors.  This is given in \cite{Freedman:1998tz} and
\begin{eqnarray}\label{propB}
G_{m \mu}( y, z; x )=\frac{\Gamma(4)}{2 \pi^2 \Gamma(2)}
\frac{z^2}{(z^2+({y}-{ x})^2)^3}\big(\delta_{m \mu }-2
\frac{(y-x)_{m} ~ (y-x)_\mu}{z^2+({y}-{x})^2}  \big),
\end{eqnarray}
where $(y-x)_4 \equiv  z$. 
To simplify the resulting expressions we take the limit in which the super gravity mode is 
located far away from the semi-classical states. 
Taking  $x\rightarrow \infty$, the bulk to boundary propagator reduces to 
\be
\label{propB1}
 G_{m \mu}(z, {y}; x ) =\frac{3}{\pi^2} \left\{ \begin{array}{ll}
         \frac{z^2}{({\vec x}^2)^3} (\delta_{m \mu}-\frac{2 x_{m}x_\mu}{{ x}^2})& \mbox{if $m=0,1,2,3$};\\
        \frac{z^2}{({ x}^2)^3}(\frac{-2 z(z-x)_\mu}{{ x}^2})\approx 0 & \mbox{if $m=4$ since $\delta_{4\mu}=0$}.\end{array} \right. 
\ee

Now let us consider any  semi-classical  solution 
which is point-like in $AdS$ and  which is either point-like or extended in the 
in the deformed sphere.  The solution can be written as 
\begin{eqnarray}\label{str-sol}
z=\frac{R}{\cosh(k \tau) },\,\,\,\,y^0=R \tanh(k \tau), \\ \nonumber
\a,\theta,\phi_1,\phi_2,\phi_3=\a,\theta,\phi_1,\phi_2,\phi_3(\sigma,\tau).
\end{eqnarray}
and $R=\frac{|x^0_1-x^0_2|}{2}$ refers to distance between the end points of the solution on the
boundary.    Note that the end points are separated only in the $0$ direction along the 
boundary. 
The world-sheet dependence on the angular co-ordinates is such that 
it satisfies the equations of motion and the Virasoro constraints. 
Substituting the solution in (\ref{str-sol}) into 
(\ref{expB1})  and using  the expression for the 
bulk to boundary propagator given in (\ref{propB1}) we obtain 
\be
\label{expB2}
\langle B_\mu( x)  \rangle=-\frac{3\sqrt{\lambda}}{ 2 \pi^3}\frac{1}{({x}^2)^3} (\delta_{m \mu}-\frac{2 x_{m}x_\mu}{{ x}^2}) \,\, I^{m}
\ee
Since the string extends only in $y^0$ and $z$ directions, 
the index $m$  for $I^m$ can be either 0 or 4. But as we have seen 
the bulk to boundary propagator  (\ref{propB1}), $G_{4\mu}=0$ in the limit $x\rightarrow \infty$.
Therefore the only non-trivial contribution is from  $m=0$ and $I^0$ is given by 
\begin{eqnarray}
\label{I}
I^{0}&=&\int d^2\s \pa_\t y^0(\pa_\t \phi^1 W_1+\pa_\t \phi^2 W_2+\pa_\t \phi^3 W_3 +i\e^{01}(\pa_\s \phi^2 Z_2+\pa_\s \phi^3 Z_3) )z^2,  \\ \nonumber
&=&\int d^2\s \pa_\t y^0 \frac{R^2}{\cosh^2{k\tau}} \, G\left(\pa_\t \phi^1+\g^2r_2^2r_3^2\sum_i^3 \pa_\t \phi^i +i\e^{01}(r_2^2\pa_\s \phi_2 - r_3^2\pa_\s \phi_3)\right). \\ \nonumber
\end{eqnarray}
Now we note that the conserved charge corresponding to 
shifts in $\phi_1$  is given by 
\begin{eqnarray}
J&=&\frac{\sqrt{\l}}{2\p}\int d\s \frac{\pa{\cal L}_{string}}{\pa(\pa_\t \phi_1)}, \\ \nonumber
&=&\frac{\sqrt{\l}}{2\p}\int d\s G\left[\pa_\t \phi^1+\g^2r_2^2r_3^2\sum_i^3 \pa_\t \phi^i +i\e^{01}(r_2^2\pa_\s \phi_2 - r_3^2\pa_\s \phi_3)\right].
\end{eqnarray}
Using this expression for the $U(1)$ R-charge,  $\ref{I}$ can be written as
\be
\label{I1}
I^{0}=\frac{2\p J}{\sqrt{\l}} \int_{-\infty}^{\infty} d\t \frac{k R^3}{\cosh^4{k \t}}=\frac{8R^3J \p}{3\l}. 
\ee
Substituting (\ref{I1}) in (\ref{expB2}) we obtain
\begin{eqnarray}\label{expB3}
\langle B_\mu( x)  \rangle=-J\frac{4 R^3}{\pi^2}\frac{1}{({\vec x}^2)^3} 
(\delta_{0 \mu}-\frac{2 x_{0}x_\mu}{{ x}^2}).
\end{eqnarray}
Now dividing the three point function given in (\ref{vector2})
by the 2 point function of ${\cal O}_\D$ and substituting  $x^0_{12}=2R$ 
we obtain 
\begin{eqnarray}
\label{vector1}
\frac{\langle j_{\mu}^R(x) \,\,{\cal O}_{\Delta}(x_1)\,\,{\bar {\cal O}}_{\Delta}(x_2) \rangle}{\langle {\cal O}_{\Delta}(x_1)\,\,{\bar {\cal O}}_{\Delta}(x_2) \rangle}=C(\lambda) (2R)^3\frac{1}{({\vec x}^2)^3} (\delta_{0 \mu}-\frac{2 x_{0}x_\mu}{{\vec x}^2}).
\end{eqnarray}
Comparing (\ref{vector1}) and (\ref{expB3}) we determine the structure constant 
for the $U(1)$ R-current with semi-classical states to be 
\be
\label{string_sc}
C(\lambda>>1)=-\frac{J}{2 \pi^2}.
\ee
Thus the structure constant evaluated in the Lunin-Maldacena 
geometry agrees with the all loop 
prediction resulting from the 
ward identity given in (\ref{C123_R}).
The same calculation can be easily 
repeated for the remaining two R-currents corresponding to the 
shifts in $\phi_2$ and $\phi_3$ resulting in the same conclusion.

\subsection{Structure constant involving rigid rotating strings}

In this section we evaluate the structure constant involving 
the supergravity mode dual to the $(k, k, k)$ chiral primary and 
a semi-classical rigid  string solution  which has equal R-charges in the 
three $U(1)$s. 
The string solution we consider is given by 
\begin{eqnarray}\label{semst}
& & x^{0} = R \tanh \kappa\tau, \qquad z =  \frac{R}{\cosh \kappa \tau}, \qquad
R= \frac{1}{2} |x_1 - x_2|, \\ \nonumber
& & r_1=r_2=r_3=\frac{1}{\sqrt{3}},\qquad \phi_1=\phi_2=i \w \t+m\s, \qquad \phi_3=i\w\t-2m\s.
\end{eqnarray}
This configuration is a solution to the world sheet equations of motion in the Lunin-Maldacena 
geometry. $m$ is an integer which refers to the winding of the string along the 
$U(1)$. For the string to be a massive state we take $m\neq 0$. 
It has equal R-charges in the three $U(1)$'s with a total R-charge given by 
\be
J=J_1+J_2+J_3=\sqrt{\l} \w.
\ee
The configuration given in (\ref{semst}) satisfies the Virasoro constraint
corresponding to world sheet momentum. 
\begin{equation}
G_{MN} \partial_{\tau} X^M\partial_{\sigma} X^N = 0, 
\end{equation}
The Virasoro constraint corresponding to the world sheet energy reduces to 
to the following constraint among the parameters of the solution. 
\be \label{virig}
\k^2-\w^2=\frac{ 6 m^2}{3+\g^2}. 
\ee
Rewriting this constraint in terms of the energy $E = \sqrt{\lambda} \kappa$ and 
R-charge $J = \sqrt{\lambda} \omega$ we obtain
\begin{equation}
E^2-J^2=\frac{\l 6 m^2}{3+\g^2} .
\end{equation}

Then the structure constant involving the rigid string and the super gravity mode 
dual to the $(k, k, k )$ chiral primary is given by evaluating the amplitude given in 
(\ref{struc1}) by substituting the classical configuration in  (\ref{semst}). 
This results in 
\be
\label{strucrigid1}
{\cal{C}}_{\rm{rigid}}=\frac{2^{-k}(k+1) \sqrt{k \l}}{8\pi N}\int d^2\s \frac{e^{- k \w \t }}{\cosh^k \k \t}\left[\frac{(\partial x)^2-(\partial z)^2}{z^2}-H^{\a \b} \partial_a X^{\a} \partial_b X^{\b} h^{ab}\right].
\ee
Note that there is no contribution from the NS-B field  fluctuations given in (\ref{amflu}) 
even though this classical configuration has world sheet $\sigma$ dependence. 
The expression in square brackets in (\ref{strucrigid1}) can simplified as follows
\begin{eqnarray}
\frac{(\partial x)^2-(\partial z)^2}{z^2}-H^{\a \b} \partial_a X^{\a} \partial_b X^{\b} h^{ab}&=&\k^2(-1 + 2\operatorname{sech}^2 \k \t )+\w^2+\frac{6(-3+\g^2) m^2}{(3+\g^2)^2} \nonumber \\  
&=&2\kappa^2\operatorname{sech}^2 \k\t-\frac{36 m^2}{(3+\g^2)^2} .
\end{eqnarray}
To obtain the second line in the above equation we have used the 
Virasoro constraint (\ref{virig}).  Substituting this result in (\ref{strucrigid1}) we obtain
\be
\label{strucrigid2}
C_{{\rm rigid}}=\frac{2^{-k} (k+1) \sqrt{k \l}}{4 N}\int^{\infty}_{-\infty}\left[\frac{2 \k^2 e^{k\w\t}}{\operatorname{cosh}^{k+2}\k \t}-\frac{36 m^2}{(3+\g^2)^2}\frac{e^{-k \w \t}}{\operatorname{cosh}^k \k \t}\right] d\t.
\ee
The integrals in the above expression  can be easily performed using 
\begin{eqnarray}
\label{rigint}
\int^{\infty}_{-\infty}\frac{e^{-k \w \t}}{\operatorname{cosh}^{k+2} \k \t}d\t &=& \frac{2^{a_1-1}}{\k}B\left(\frac{a_1+b}{2},\frac{a_1-b}{2}\right), \nonumber  \\ 
 \int^{\infty}_{-\infty}\frac{e^{-k \w \t}}{\operatorname{cosh}^{k} \k \t}d\t&=&\frac{2^{a_2-1}}{\k}B\left(\frac{a_2+b}{2},\frac{a_2-b}{2}\right),
\end{eqnarray}
where  $B(x, y)$ is the Beta function and 
\begin{equation}
a_1=k+2, \quad a_2=k,  \quad b=\frac{k \w}{\k}=\frac{k J}{E}.
\end{equation} 
Now substituting  the results for the integrals given in  (\ref{rigint}) into  (\ref{strucrigid2}), we obtain
\be
\label{strucrigid3}
C_{rigid}=\frac{ (k+1) \sqrt{k \l}}{4 \k N}\left[4\k^2 B\left(\frac{a_1+b}{2},\frac{a_1-b}{2}\right)-\frac{18 m^2}{(3+\g^2)^2} B\left(\frac{a_2+b}{2},\frac{a_2-b}{2}\right)\right].
\ee
We then use 
the following  property of the beta function
\be
B(x+1,y+1)=\frac{xy}{(x+y+1)(x+y)}B(x,y)
\ee
to further simplify (\ref{strucrigid3}). This results in 
\be
\label{strucrigid4}
C_{\rm rigid} = 
\frac{3 m^2 (\g^2 k-3)\sqrt{k\l}}{2\k N(3+\g^2)^2}B\left(\frac{k(\k+\w)}{2\k},\frac{k(\k-\w)}{2\k}\right).
\ee
Finally re-writing this structure constant in terms of the charges $E$ and $J$ of the rigid string
we obtain 
\begin{equation}
\label{strucrigid5}
C_{\rm rigid} = \frac{ ( E^2 - J^2)( \gamma^2 k -3) \sqrt{k}}{ 4E ( 3 + \gamma^2)} 
B \left( \frac{ k (E + J)}{2E}, \frac{ k ( E-J)}{ 2E} \right).
\end{equation}
Note that this strong coupling result vanishes for $\gamma^2 k =3$, it will be interesting 
to verify this directly in the field theory.

\section{Conclusions}

We have studied the structure constants of the ${\cal N}=1$ beta deformed theory 
both perturbatively at one loop and in strong coupling using the Lunin-Maldacena
geometry. 
We have shown that the  three point function of the chiral primaries 
with equal $U(1)$ charges along the three Cartan directions are not renormalized 
at one loop and their value is 
independent of the deformation $\beta$ . Therefore it is the same as that in the ${\cal N}=4$ theory. 
We have observed the same
 behaviour of the these three point functions  at strong
coupling in the Lunin-Maldacena geometry.  This suggests that these
three point functions are not renormalized in the beta deformed theory. 
It will be interesting to see if the methods developed to prove non-renormalization
of 3 point functions of chiral primaries in the ${\cal N}=4$ theory  
\cite{Howe:1998zi,Basu:2004nt,Baggio:2012rr} can be carried over 
for these chiral primaries of the beta deformed theory. 

The three point function of the chiral primaries in the Lunin-Maldacena background
were evaluated in the limit when two of the chiral primaries 
have large R-charges and are semi-classical
using the methods of \cite{Janik:2010gc,Zarembo:2010rr}. It will be interesting  to evaluate them for 
of arbitrary R-charges following  \cite{Lee:1998bxa}. 
The supergravity modes  in the 
Lunin-Maldacena geometry dual to the chiral primary with $U(1)$ charges $(k, k, k )$ as 
well that of the R-currents were constructed by TsT transformation of the corresponding 
mode in the $AdS_5\times S^5$ background. It will interesting to construct 
the supergravity modes corresponding the other chiral primaries with 
$U(1)$ charges $(k, 0, 0)$, $(0, k, 0)$ and $(0, 0, k)$ in the Lunin-Maldacena background
to verify similar non-renormalization properties of their three point functions. 

As the beta deformed theory is the only theory with ${\cal N}=1$ supersymmetry 
which is known to be integrable in the planar limit
it will be useful to apply all the methods developed for studying the ${\cal N}=4$ 
theory which relied on its integrability. 
Most likely this general direction will yield useful results just as this study of 
the structure constants has  demonstrated.

\acknowledgments

J.R.D  thanks the hospitality of the string theory group at Humbolt University, Berlin 
and for an opportunity to present this work.  We thank Harald  Dorn,
George Jorjadze,  Jan Plefka, Christoph Sieg and Mathias Staudacher
for useful and stimulating discussions. 
J.R.D thanks  the organizers of the 7th Regional meeting in string theory, Crete
for hospitality and a  stimulating meeting. 
We also thank  Kyriakos Papadodimas and Konstantinos Zoubos for discussions
on non-renormalization theorems for  3 point functions of chiral primaries.  
The work of  J.R.D is partially supported by
the Ramanujan fellowship DST-SR/S2/RJN-59/2009, the work of A.S is supported by  a
CSIR fellowship (File no: 09/079(2372)/2010-EMR-I).

\begin{appendix}

\section{Anomalous dimensions }

In this section we review the evaluation of the anomalous dimension Hamiltonian 
for single trace operators constructed out of the three complex scalars in the 
beta deformed theory. 
The two point function of two of such operators $O_\alpha, O_\beta$ at one loop 
is given by 
 \begin{equation}\label{defanh}
  \langle O_\alpha (x_1)  O_\beta (x_3) \rangle = 
  \frac{1}{|x_1 - x_3|^{2\Delta_\alpha} }\left( h_{\alpha\beta} + \lambda g_{\alpha\beta}
  - \lambda \gamma_{\alpha\beta} \ln ( ( x_1 -x_3) ^2 \Lambda^2 )  \right).
  \end{equation}
  The anomalous dimension  Hamiltonian 
  at one loop is determined  by the terms proportional to the 
  logarithm   in the above expression. 
  By simple large $N$ counting it can be seen that only interactions between nearest 
  neighbour letters contribute at one loop. 
  Therefore to characterize the anomalous dimension Hamiltonian it is sufficient to 
  extract the log terms  in the interactions given in figures \ref{hs}, \ref{x1}, \ref{x2}
  where $A, B$ are nearest neighbour letters in say operator $O_\alpha(x_1)$ and 
  $B, C$ are nearest neighbour letters in operator $O_\beta(x_1)$. 
  Figure \ref{hs}  contains the gauge exchange and the self energy interactions, 
 figure \ref{x1} contains the quartic interactions from the D-type quartic interactions 
 and \ref{x2} contains the possible quartic interactions from the F-type quartic
 interactions. 
 We will now discuss the various possibilities the letters $A, B, C,D$ can take 
 in the same order as we discussed for the case of the structure constants. 
 We will see that the coefficient of the log term is proportional to the contribution 
 of the structure constant. 
 By large $N$ counting it is easily seen that the coefficients of the 
 log term is proportional to $\lambda$.  We will suppress this dependence in the 
 discussion below. 
 
 \subsection*{(i) $A =B=Z_i$ and $C= D=\bar Z_i$, $i = 1, 2, 3$ }
 
 We consider the contribution to the log term with both the letters $A, B $ being
 $Z_i$ and the letters $C, D$ being $\bar Z_i$. $i$ can take any value
 $i = 1, 2, 3$. 
 Here instead of explicitly evaluating the contribution of the self energy diagrams
 we will use the fact that this combination is letters occurs in the evaluation 
 of the one loop contribution of the anomalous dimension of a chiral primary
 for eg. the operator ${\rm Tr}( Z_1^k) $.  Thus the sum  of the 
 coefficients of the  terms proportional to the logarithim must vanish in this case. 
 This will determine the contribution of the self energy diagrams in terms of the 
 others. 
The quartic term which contributes in this situation arises from 
diagrams in figure \ref{q3} and figure \ref{q4}. Writing out the contributions of all these diagrams
we obtain
\begin{eqnarray}
-\tilde \Delta( {\bf i} ) &=& - \frac{1}{2} \lim_{x_1\rightarrow x_2, x_3 \rightarrow x_4} H(x_1, x_2, x_3) 
+ S(x_1, x_3)    \\ \nonumber
& & - \frac{2}{4} \lim_{x_1\rightarrow x_2, x_3 \rightarrow x_4}  \frac{\phi(r, s) }{x_{13}^2 x_{24}^2}.
\end{eqnarray} 
The limits $x_1\rightarrow x_2$ is taken by setting $x_1 - x_2 = \epsilon$ and then
taking $\epsilon \rightarrow 0$.  The same procedure is adopted to take the limit
$x_3\rightarrow x_4$.  
The reason for the factors of $1/2$  and signs are the same as that explained in the 
evaluation of the structure constant contribution. 
 We have inserted the negative sign  in front of $-\tilde \Delta$ because  from 
 the equation in (\ref{defanh}) 
the anomalous dimension Hamiltonian is negative the coefficient of the logarithm. 
We now extract the coefficients proportional to the logarithm in each of the terms above 
 to obtain the contributions to the anomalous dimension Hamiltonian
 \begin{equation}
 -\Delta({\bf i} ) = H + S - \frac{1}{2} X .
  \end{equation}
  where $H$ is the coefficient of the term proportional  to the gauge exchange diagram, 
  $S$ is the coefficient from the self energy diagram and $X$ is the coefficient from 
  $\phi(r, s)$, the quartic diagram. 
Since $\Delta({\bf i} )$ vanishes we obtain the relation
\begin{equation}
H+ S = \frac{1}{2} X.
\end{equation}
From the expansion of the function $\phi(r, s)$ on taking the limits 
$x_1\rightarrow x_2, x_3 \rightarrow x_4$ it can be seen that 
\begin{equation}
X=2.
\end{equation} 
To summarize:  for this combination of letters we have 
\begin{equation}
\Delta({\bf i} ) =0.
\end{equation}
Note that the contribution to the anomalous dimension Hamiltonian from diagram 
with $A = B = \bar Z_i$ and $C= D =  Z_i$, that is with the $Z$'s interchanged 
with the corresponding $\bar Z$ also vanishes. 

\subsection*{ (ii) $A= D = Z_i$ and $B=C= \bar Z_i$,  $i = 1, 2, 3$}

The diagrams which contribute in this case are the gauge exchange, the self energies, 
and the quartic diagrams given in (\ref{q1}) and (\ref{q2}). 
The contribution to the anomalous dimensions is given by 
\begin{eqnarray}
-\Delta({\bf ii} ) &=& H+ S + \frac{1}{2} X, \\ \nonumber
 &=& X = 2.
 \end{eqnarray}
 An identical contribution arises from the complex conjugated diagram. 
 
 \subsection*{(iii)  $A = Z_i, B= Z_j, C= \bar Z_i, D= \bar Z_j$ and $i\neq j$, $i = 1, 2 , 3$}
 
 Here the gauge exchange and  the self energies contribute. 
 The quartic interaction both from the D-term and the F-term contributes.
 The D-term contribution arise from diagrams in (\ref{q3}) and (\ref{q4}). 
 The F-term contribution arise from diagrams of the type in  (\ref{q5}) or
 (\ref{q6}).  Putting these contributions together we obtain
 \begin{eqnarray}
 -\Delta({\bf iii} ) &=& = H+ S - \frac{1}{2} X + X , \\ \nonumber
 &=& X = 2.
 \end{eqnarray}
 Here again the complex conjugate of this diagram yields the same contribution 
 to the anomalous dimension Hamiltonian. 
 
 \subsection*{ (iv) $ A =Z_i, B= \bar Z_j, C = \bar Z_i, D = \bar Z_j$ and $i\neq j$, $i =1, 2, 3$}
 
 Here the gauge exchange and the self energy diagrams contribute.
 The only contribution for this class of diagrams  from the quartic interaction in 
 (\ref{q2}). This diagram has to be counted with a factor of $2$ since there is two possible 
 ways to Wick contract.  This results in 
 \begin{eqnarray}
 -\Delta({\bf iv} ) &=&  H+S +  \frac{2}{4} X , \\ \nonumber
 &=& X = 2.
 \end{eqnarray}
The complex conjugate diagram also gives the same result. 
 
 \subsection*{(v) $ A= Z_i, B = \bar Z_i, C= \bar Z_j, D = \bar Z_j$ and $i\neq j$, $i =1, 2, 3$}
 
Here the only diagram which contributes is from the D-type interaction in 
(\ref{q1}) with a factor of $2$ due to the two possible ways of Wick contraction. 
There is no contribution from the gauge exchange or the self energy diagrams. 
The contribution to the anomalous dimension Hamiltonian is therefore
\begin{equation}
-\Delta({\bf v} ) = \frac{2}{4} X = 1.
\end{equation}
Here again the complex conjugate diagram yields the same result. 

\subsection*{(vi) $A = Z_i, B= \bar Z_i , C= Z_j, D= \bar Z_j$ and $i\neq j$, $i = 1, 2, 3$}

Here quartic interaction from D-type terms in (\ref{q3}) and (\ref{q4})  contribute. 
Then one gets a contribution either from the F-type terms in 
(\ref{q5}) or (\ref{q6}). 
There is no contribution from the self energy or the gauge exchange. 
Putting this together results in 
\begin{equation}
-\Delta({\bf vi} ) = -\frac{2}{4} X + X = \frac{1}{2} X = 1.
\end{equation}
The same contribution results from a complex conjugate of this diagram. 

\subsection*{(vii) $A=Z_1, B= Z_2, C= \bar Z_2, D = \bar Z_1$ and cyclic}

The only contribution arises from the F-type interaction in (\ref{q7}). 
There is no contribution from the self energy or the gauge exchange. 
This results in 
\begin{equation}
-\Delta({\bf vii} ) = - e^{i\beta} X = - 2e^{i\beta} .
\end{equation}
The same contribution is obtained in all diagrams obtained by the 
cyclic replacement $Z_1 \rightarrow Z_2\rightarrow Z_3\rightarrow Z_1$. 
Complex conjugate of this diagram also yields the same result. 

\subsection*{(viii) $A = Z_2, B=Z_1, C= Z_1, D= \bar Z_2$  and cyclic}

There is no contribution from the gauge exchange and the self energy. 
The only contribution is from the F-type diagram in (\ref{q8}). 
Therefore we have
\begin{equation}
-\Delta({\bf viii} ) = - e^{-i\beta} X = - 2 e^{-i\beta}.
\end{equation}
All diagrams obtained by the cyclic replacements as well as conjugation
yields the same result for the anomalous dimensions. 

\subsection*{(ix) $A=Z_1, B= \bar Z_2, C= Z_2, D= \bar Z_1$ and cyclic}

Here the only contribution is from the F-type diagram in (\ref{q8}). 
This gives
\begin{equation}
-\Delta({\bf ix} ) = - e^{-i\beta} X = - 2 e^{-i\beta}.
\end{equation}
Again all diagrams obtained by the cyclic replacements as well as conjugation
yields the same result for the anomalous dimensions. 

\subsection*{(x) $A= Z_2, B= \bar Z_1, C= Z_1, D= \bar Z_2$  and cyclic}

The only contribution is from the F-type diagram of the type in (\ref{q7})
which gives the following 
\begin{equation}
-\Delta({\bf x} ) = - e^{i\beta} X =- 2 e^{i\beta}.
\end{equation}
All cyclically related diagrams and those obtained by conjugation give 
the same result for the anomalous dimensions.

\subsection*{$SU(3)$ sub-sector}

The $SU(3)$ sub-sector is defined by single trace gauge invariant operators made up of only the holomorphic combinations of the letters $Z_1, Z_2, Z_3$. 
One can easily examine the above calculations for the $SU(3)$ subsector and show that 
the  contribution for anomalous dimension Hamiltonian can be written in terms of 
the following Hamiltonian \cite{Minahan:2011bi}
\be
{\cal D}_{ij}=\frac{\lambda}{4\pi}({\cal H}+{\cal H}')_{ij},
\ee
where
\be
{\cal H}_{ij}=\frac{1}{2}(\lm^3_i\lm^3_j+\lm^8_i\lm^8_j-\frac{4}{3}I_{ij})
\ee
and
\be
{\cal H}'_{ij}=e^{i \beta}(\lm^{+1}_i\lm^{-1}_j+\lm^{-2}_i\lm^{+2}_j+\lm^{+3}_i\lm^{-3}_j)+e^{-i\beta}(\lm^{-1}_i\lm^{+1}_j+\lm^{+2}_i\lm^{-2}_j+\lm^{-3}_i\lm^{+3}_j),
\ee
where $\lm^a$, $a=1...,8$ are $3\times 3$ Gell-Mann matrices. 
These  act on the letters of the single gauge invariant operators  at nearest neighbours $i$ and $j$. 
The raising and lowering operators are defined by 
\begin{equation}
 \lm^{\pm 3}=\frac{1}{2}(\lm^1\pm i\lm^2),  \qquad 
  \lm^{\pm 2}=\frac{1}{2}(\lm^4\pm i\lm^5), \qquad  \lm^{\pm 1}=\frac{1}{2}(\lm^6\pm i\lm^7),
  \end{equation}
 and $I$ is the identity operator. 

\subsection*{Twisted Hamiltonian}

To write down the anomalous dimension Hamiltonian for the full sector in a compact form
it is convenient to realize the Hamiltonian as twisted version of the one loop 
Hamiltonian of ${\cal N}=4$ Yang-Mills  \cite{Beisert:2005if}. 
For this we  define a charge  vector ${\bf q}$ for the fields 
$Z_i$ and the  charge vector $\bar{\bf q}$ for the complex conjugates
$\bar Z_i$ according to table \ref{t1}. 
\begin{center}
\begin{table}
\centering
    \begin{tabular}{ | l | l | l | l | l | l | l | l|}
    \hline
 & $Z_1$ & $Z_2$ & $Z_3$ &   & $\bar{Z_1}$ & $\bar{Z_2}$ & $\bar{Z_3}$ \\ \hline
$q_\alpha^1$ & 1 & 0 & 0 & $\bar{q}_\alpha^1$ & $-1$ & $0$ & $0$  \\ 
$q_\alpha^2$ & 0 & 1 & 0 & $\bar{q}_\alpha^2$ & $0$ & $-1$ & $0$  \\ 
$q_\alpha^3$ & 0 & 0 & 1 & $\bar{q}_\alpha^3$ & $0$ & $0$ & $-1$  \\ \hline
    \end{tabular}
    \caption{ $U(1)$ charges of scalar fields}
    \label{t1}
\end{table}
\end{center}
We then define the {\bf C}-product between two charge vectors  by 
\be\label{cprod}
{\bf{q}}_\alpha \times {\bf{q}}_\beta={\bf{q}}_\alpha^T {\bf{C}} {\bf{q}}_\beta=
C_{ab}q^{a}_\alpha q^{b}_\beta,
\ee
where $\alpha, \beta$ labels the vectors and 
 $\bf{C}$ is defined as
\be
\label{C}
{\bf{C}}=\left( \begin{array}{ccc}
   0 & -\b & \b \\
 \b & 0 & -\b \\
  -\b & \b & 0 \\
\end{array}
  \right).
\ee
In (\ref{cprod})  any of the vectors can also correspond to the charges of the conjugate field. 
Let the anomalous dimension Hamiltonian of ${\cal N}=4$ Yang-Mills be denoted by 
$ {\cal{H}}^{\alpha\beta}_{\gamma\delta}$. Here $\alpha, \beta$ refer to the nearest
neighbour letters in one operator, and $\gamma, \delta$ refers to the nearest 
neighbour letters in the second operator. 
Then the anomalous dimension Hamiltonian of the beta deformed theory is 
related to the anomalous dimension Hamiltonian of ${\cal N}=4$ Yang-Mills by 
  \be
\tilde {\cal{H}}^{\alpha\beta}_{\gamma\delta}=
e^{i ({\bf q}_\alpha \times {\bf {q}}_\beta-{\bf q}_\gamma \times {\bf {q}}_\delta)/2} 
{\cal{H}}^{\alpha\beta}_{\gamma\delta}.
\ee
Here $\tilde{\cal H} $ refers to the anomalous dimension Hamiltonian of the deformed theory. 
In using the above relation, one has to substitute the charge vector ${\bar q}$
if any of the letters is the complex conjugate. 
It can be easily verified for all the combination of letters that $\tilde {\cal H}$ 
reproduces the values found by the explicit calculation.

\end{appendix}

\bibliographystyle{JHEP}
\bibliography{structure}

\providecommand{\href}[2]{#2}\begingroup\raggedright\begin{thebibliography}{10}

\bibitem{Beisert:2010jr}
N.~Beisert, C.~Ahn, L.~F. Alday, Z.~Bajnok, J.~M. Drummond, et~al., {\it
  {Review of AdS/CFT Integrability: An Overview}},  {\em Lett.Math.Phys.} {\bf
  99} (2012) 3--32, [\href{http://xxx.lanl.gov/abs/1012.3982}{{\tt
  arXiv:1012.3982}}].

\bibitem{Okuyama:2004bd}
K.~Okuyama and L.-S. Tseng, {\it {Three-point functions in N = 4 SYM theory at
  one-loop}},  {\em JHEP} {\bf 0408} (2004) 055,
  [\href{http://xxx.lanl.gov/abs/hep-th/0404190}{{\tt hep-th/0404190}}].

\bibitem{Roiban:2004va}
R.~Roiban and A.~Volovich, {\it {Yang-Mills correlation functions from
  integrable spin chains}},  {\em JHEP} {\bf 0409} (2004) 032,
  [\href{http://xxx.lanl.gov/abs/hep-th/0407140}{{\tt hep-th/0407140}}].

\bibitem{Alday:2005nd}
L.~F. Alday, J.~R. David, E.~Gava, and K.~Narain, {\it {Structure constants of
  planar N = 4 Yang Mills at one loop}},  {\em JHEP} {\bf 0509} (2005) 070,
  [\href{http://xxx.lanl.gov/abs/hep-th/0502186}{{\tt hep-th/0502186}}].

\bibitem{Lee:1998bxa}
S.~Lee, S.~Minwalla, M.~Rangamani, and N.~Seiberg, {\it {Three point functions
  of chiral operators in D = 4, N=4 SYM at large N}},  {\em
  Adv.Theor.Math.Phys.} {\bf 2} (1998) 697--718,
  [\href{http://xxx.lanl.gov/abs/hep-th/9806074}{{\tt hep-th/9806074}}].

\bibitem{Janik:2010gc}
R.~A. Janik, P.~Surowka, and A.~Wereszczynski, {\it {On correlation functions
  of operators dual to classical spinning string states}},  {\em JHEP} {\bf
  1005} (2010) 030, [\href{http://xxx.lanl.gov/abs/1002.4613}{{\tt
  arXiv:1002.4613}}].

\bibitem{Buchbinder:2010vw}
E.~Buchbinder and A.~Tseytlin, {\it {On semiclassical approximation for
  correlators of closed string vertex operators in AdS/CFT}},  {\em JHEP} {\bf
  1008} (2010) 057, [\href{http://xxx.lanl.gov/abs/1005.4516}{{\tt
  arXiv:1005.4516}}].

\bibitem{Zarembo:2010rr}
K.~Zarembo, {\it {Holographic three-point functions of semiclassical states}},
  {\em JHEP} {\bf 1009} (2010) 030,
  [\href{http://xxx.lanl.gov/abs/1008.1059}{{\tt arXiv:1008.1059}}].

\bibitem{Costa:2010rz}
M.~S. Costa, R.~Monteiro, J.~E. Santos, and D.~Zoakos, {\it {On three-point
  correlation functions in the gauge/gravity duality}},  {\em JHEP} {\bf 1011}
  (2010) 141, [\href{http://xxx.lanl.gov/abs/1008.1070}{{\tt
  arXiv:1008.1070}}].

\bibitem{Roiban:2010fe}
R.~Roiban and A.~Tseytlin, {\it {On semiclassical computation of 3-point
  functions of closed string vertex operators in $AdS_5 x S^5$}},  {\em
  Phys.Rev.} {\bf D82} (2010) 106011,
  [\href{http://xxx.lanl.gov/abs/1008.4921}{{\tt arXiv:1008.4921}}].

\bibitem{Ryang:2010bn}
S.~Ryang, {\it {Correlators of Vertex Operators for Circular Strings with
  Winding Numbers in AdS5xS5}},  {\em JHEP} {\bf 1101} (2011) 092,
  [\href{http://xxx.lanl.gov/abs/1011.3573}{{\tt arXiv:1011.3573}}].

\bibitem{Escobedo:2010xs}
J.~Escobedo, N.~Gromov, A.~Sever, and P.~Vieira, {\it {Tailoring Three-Point
  Functions and Integrability}},  {\em JHEP} {\bf 1109} (2011) 028,
  [\href{http://xxx.lanl.gov/abs/1012.2475}{{\tt arXiv:1012.2475}}].

\bibitem{Escobedo:2011xw}
J.~Escobedo, N.~Gromov, A.~Sever, and P.~Vieira, {\it {Tailoring Three-Point
  Functions and Integrability II. Weak/strong coupling match}},  {\em JHEP}
  {\bf 1109} (2011) 029, [\href{http://xxx.lanl.gov/abs/1104.5501}{{\tt
  arXiv:1104.5501}}].

\bibitem{Foda:2011rr}
O.~Foda, {\it {N=4 SYM structure constants as determinants}},  {\em JHEP} {\bf
  1203} (2012) 096, [\href{http://xxx.lanl.gov/abs/1111.4663}{{\tt
  arXiv:1111.4663}}].

\bibitem{Gromov:2012uv}
N.~Gromov and P.~Vieira, {\it {Tailoring Three-Point Functions and
  Integrability IV. Theta-morphism}},
  \href{http://xxx.lanl.gov/abs/1205.5288}{{\tt arXiv:1205.5288}}.

\bibitem{Bissi:2012vx}
A.~Bissi, G.~Grignani, and A.~Zayakin, {\it {The SO(6) Scalar Product and
  Three-Point Functions from Integrability}},
  \href{http://xxx.lanl.gov/abs/1208.0100}{{\tt arXiv:1208.0100}}.

\bibitem{Elmetti:2006gr}
F.~Elmetti, A.~Mauri, S.~Penati, and A.~Santambrogio, {\it {Conformal
  invariance of the planar beta-deformed N=4 SYM theory requires beta real}},
  {\em JHEP} {\bf 0701} (2007) 026,
  [\href{http://xxx.lanl.gov/abs/hep-th/0606125}{{\tt hep-th/0606125}}].

\bibitem{Elmetti:2007up}
F.~Elmetti, A.~Mauri, S.~Penati, A.~Santambrogio, and D.~Zanon, {\it {Real
  versus complex beta-deformation of the N=4 planar super Yang-Mills theory}},
  {\em JHEP} {\bf 0710} (2007) 102,
  [\href{http://xxx.lanl.gov/abs/0705.1483}{{\tt arXiv:0705.1483}}].

\bibitem{Lunin:2005jy}
O.~Lunin and J.~M. Maldacena, {\it {Deforming field theories with U(1) x U(1)
  global symmetry and their gravity duals}},  {\em JHEP} {\bf 0505} (2005) 033,
  [\href{http://xxx.lanl.gov/abs/hep-th/0502086}{{\tt hep-th/0502086}}].

\bibitem{Frolov:2005iq}
S.~Frolov, R.~Roiban, and A.~A. Tseytlin, {\it {Gauge-string duality for
  (non)supersymmetric deformations of N=4 super Yang-Mills theory}},  {\em
  Nucl.Phys.} {\bf B731} (2005) 1--44,
  [\href{http://xxx.lanl.gov/abs/hep-th/0507021}{{\tt hep-th/0507021}}].

\bibitem{Beisert:2005if}
N.~Beisert and R.~Roiban, {\it {Beauty and the twist: The Bethe ansatz for
  twisted N=4 SYM}},  {\em JHEP} {\bf 0508} (2005) 039,
  [\href{http://xxx.lanl.gov/abs/hep-th/0505187}{{\tt hep-th/0505187}}].

\bibitem{Gromov:2010dy}
N.~Gromov and F.~Levkovich-Maslyuk, {\it {Y-system and $\beta$-deformed N=4
  Super-Yang-Mills}},  {\em J.Phys.} {\bf A44} (2011) 015402,
  [\href{http://xxx.lanl.gov/abs/1006.5438}{{\tt arXiv:1006.5438}}].

\bibitem{Fiamberti:2008sm}
F.~Fiamberti, A.~Santambrogio, C.~Sieg, and D.~Zanon, {\it {Finite-size effects
  in the superconformal beta-deformed N=4 SYM}},  {\em JHEP} {\bf 0808} (2008)
  057, [\href{http://xxx.lanl.gov/abs/0806.2103}{{\tt arXiv:0806.2103}}].

\bibitem{Arutyunov:2010gu}
G.~Arutyunov, M.~de~Leeuw, and S.~J. van Tongeren, {\it {Twisting the Mirror
  TBA}},  {\em JHEP} {\bf 1102} (2011) 025,
  [\href{http://xxx.lanl.gov/abs/1009.4118}{{\tt arXiv:1009.4118}}].

\bibitem{Ahn:2011xq}
C.~Ahn, Z.~Bajnok, D.~Bombardelli, and R.~I. Nepomechie, {\it {TBA, NLO Luscher
  correction, and double wrapping in twisted AdS/CFT}},  {\em JHEP} {\bf 1112}
  (2011) 059, [\href{http://xxx.lanl.gov/abs/1108.4914}{{\tt
  arXiv:1108.4914}}].

\bibitem{Berenstein:2000hy}
D.~Berenstein and R.~G. Leigh, {\it {Discrete torsion, AdS / CFT and duality}},
   {\em JHEP} {\bf 0001} (2000) 038,
  [\href{http://xxx.lanl.gov/abs/hep-th/0001055}{{\tt hep-th/0001055}}].

\bibitem{Berenstein:2000ux}
D.~Berenstein, V.~Jejjala, and R.~G. Leigh, {\it {Marginal and relevant
  deformations of N=4 field theories and noncommutative moduli spaces of
  vacua}},  {\em Nucl.Phys.} {\bf B589} (2000) 196--248,
  [\href{http://xxx.lanl.gov/abs/hep-th/0005087}{{\tt hep-th/0005087}}].

\bibitem{Freedman:2005cg}
D.~Z. Freedman and U.~Gursoy, {\it {Comments on the beta-deformed N=4 SYM
  theory}},  {\em JHEP} {\bf 0511} (2005) 042,
  [\href{http://xxx.lanl.gov/abs/hep-th/0506128}{{\tt hep-th/0506128}}].

\bibitem{Penati:2005hp}
S.~Penati, A.~Santambrogio, and D.~Zanon, {\it {Two-point correlators in the
  beta-deformed N=4 SYM at the next-to-leading order}},  {\em JHEP} {\bf 0510}
  (2005) 023, [\href{http://xxx.lanl.gov/abs/hep-th/0506150}{{\tt
  hep-th/0506150}}].

\bibitem{Mauri:2005pa}
A.~Mauri, S.~Penati, A.~Santambrogio, and D.~Zanon, {\it {Exact results in
  planar N=1 superconformal Yang-Mills theory}},  {\em JHEP} {\bf 0511} (2005)
  024, [\href{http://xxx.lanl.gov/abs/hep-th/0507282}{{\tt hep-th/0507282}}].

\bibitem{Georgiou:2013ff}
G.~Georgiou, B.-H. Lee, and C.~Park, {\it {Correlators of massive string states
  with conserved currents}},  {\em JHEP} {\bf 1303} (2013) 167,
  [\href{http://xxx.lanl.gov/abs/1301.5092}{{\tt arXiv:1301.5092}}].

\bibitem{Leigh:1995ep}
R.~G. Leigh and M.~J. Strassler, {\it {Exactly marginal operators and duality
  in four-dimensional N=1 supersymmetric gauge theory}},  {\em Nucl.Phys.} {\bf
  B447} (1995) 95--136, [\href{http://xxx.lanl.gov/abs/hep-th/9503121}{{\tt
  hep-th/9503121}}].

\bibitem{Fokken:2013aea}
J.~Fokken, C.~Sieg, and M.~Wilhelm, {\it {Non-conformality of
  {$\gamma_i$}-deformed N=4 SYM theory}},
  \href{http://xxx.lanl.gov/abs/1308.4420}{{\tt arXiv:1308.4420}}.

\bibitem{Georgiou:2009tp}
G.~Georgiou, V.~L. Gili, and R.~Russo, {\it {Operator mixing and three-point
  functions in N=4 SYM}},  {\em JHEP} {\bf 0910} (2009) 009,
  [\href{http://xxx.lanl.gov/abs/0907.1567}{{\tt arXiv:0907.1567}}].

\bibitem{Georgiou:2012zj}
G.~Georgiou, V.~Gili, A.~Grossardt, and J.~Plefka, {\it {Three-point functions
  in planar N=4 super Yang-Mills Theory for scalar operators up to length five
  at the one-loop order}},  {\em JHEP} {\bf 1204} (2012) 038,
  [\href{http://xxx.lanl.gov/abs/1201.0992}{{\tt arXiv:1201.0992}}].

\bibitem{Plefka:2012rd}
J.~Plefka and K.~Wiegandt, {\it {Three-Point Functions of Twist-Two Operators
  in N=4 SYM at One Loop}},  {\em JHEP} {\bf 1210} (2012) 177,
  [\href{http://xxx.lanl.gov/abs/1207.4784}{{\tt arXiv:1207.4784}}].

\bibitem{Howe:1998zi}
P.~S. Howe, E.~Sokatchev, and P.~C. West, {\it {Three point functions in N=4
  Yang-Mills}},  {\em Phys.Lett.} {\bf B444} (1998) 341--351,
  [\href{http://xxx.lanl.gov/abs/hep-th/9808162}{{\tt hep-th/9808162}}].

\bibitem{Basu:2004nt}
A.~Basu, M.~B. Green, and S.~Sethi, {\it {Some systematics of the coupling
  constant dependence of N=4 Yang-Mills}},  {\em JHEP} {\bf 0409} (2004) 045,
  [\href{http://xxx.lanl.gov/abs/hep-th/0406231}{{\tt hep-th/0406231}}].

\bibitem{Baggio:2012rr}
M.~Baggio, J.~de~Boer, and K.~Papadodimas, {\it {A non-renormalization theorem
  for chiral primary 3-point functions}},  {\em JHEP} {\bf 1207} (2012) 137,
  [\href{http://xxx.lanl.gov/abs/1203.1036}{{\tt arXiv:1203.1036}}].

\bibitem{Madhu:2007ew}
K.~Madhu and S.~Govindarajan, {\it {Chiral primaries in the Leigh-Strassler
  deformed N=4 SYM: A Perturbative study}},  {\em JHEP} {\bf 0705} (2007) 038,
  [\href{http://xxx.lanl.gov/abs/hep-th/0703020}{{\tt hep-th/0703020}}].

\bibitem{Ahn:2011dq}
C.~Ahn and P.~Bozhilov, {\it {Three-point Correlation Function of Giant Magnons
  in the Lunin-Maldacena background}},  {\em Phys.Rev.} {\bf D84} (2011)
  126011, [\href{http://xxx.lanl.gov/abs/1106.5656}{{\tt arXiv:1106.5656}}].

\bibitem{Alizadeh:2011yt}
D.~Arnaudov and R.~Rashkov, {\it {Three-point correlators: Examples from
  Lunin-Maldacena background}},  {\em Phys.Rev.} {\bf D84} (2011) 086009,
  [\href{http://xxx.lanl.gov/abs/1106.4298}{{\tt arXiv:1106.4298}}].

\bibitem{Bozhilov:2013bya}
P.~Bozhilov, {\it {Leading finite-size effects on some three-point correlators
  in TsT-deformed {$AdS_5 x S^5$}}},
  \href{http://xxx.lanl.gov/abs/1304.2139}{{\tt arXiv:1304.2139}}.

\bibitem{Frolov:2005dj}
S.~Frolov, {\it {Lax pair for strings in Lunin-Maldacena background}},  {\em
  JHEP} {\bf 0505} (2005) 069,
  [\href{http://xxx.lanl.gov/abs/hep-th/0503201}{{\tt hep-th/0503201}}].

\bibitem{Meessen:1998qm}
P.~Meessen and T.~Ortin, {\it {An Sl(2,Z) multiplet of nine-dimensional type II
  supergravity theories}},  {\em Nucl.Phys.} {\bf B541} (1999) 195--245,
  [\href{http://xxx.lanl.gov/abs/hep-th/9806120}{{\tt hep-th/9806120}}].

\bibitem{Kim:1985ez}
H.~Kim, L.~Romans, and P.~van Nieuwenhuizen, {\it {The Mass Spectrum of Chiral
  N=2 D=10 Supergravity on S**5}},  {\em Phys.Rev.} {\bf D32} (1985) 389.

\bibitem{Freedman:1998tz}
D.~Z. Freedman, S.~D. Mathur, A.~Matusis, and L.~Rastelli, {\it {Correlation
  functions in the CFT(d) / AdS(d+1) correspondence}},  {\em Nucl.Phys.} {\bf
  B546} (1999) 96--118, [\href{http://xxx.lanl.gov/abs/hep-th/9804058}{{\tt
  hep-th/9804058}}].

\bibitem{Minahan:2011bi}
J.~Minahan and C.~Sieg, {\it {Four-Loop Anomalous Dimensions in Leigh-Strassler
  Deformations}},  {\em J.Phys.} {\bf A45} (2012) 305401,
  [\href{http://xxx.lanl.gov/abs/1112.4787}{{\tt arXiv:1112.4787}}].

\end{thebibliography}\endgroup

\end{document}